\documentclass[11pt]{article}

\usepackage{bbm}        
\usepackage{amsthm} 
\usepackage{a4wide}
\usepackage{braket}
\usepackage{float}

\usepackage{graphics}
\usepackage[pdftex]{epsfig}
\usepackage{epstopdf}

\newcommand{\I}{\mathbbm 1}

\graphicspath{{./Figures/}} 

\title{Oblivious transfer based \\on single-qubit rotations}
\author{Jo\~ao Rodrigues\textsuperscript{1,2}, Paulo Mateus\textsuperscript{1,2},\\ Nikola Paunkovi\'{c}\textsuperscript{1,2} and Andr\'e Souto\textsuperscript{3,4}\thanks{email: ansouto@fc.ul.pt}}
\date{}

\begin{document}

%André
\def\ds{\displaystyle}
\newtheorem{theo}{Theorem}%[section]
\newtheorem{lemma}[theo]{Lemma}
\newtheorem{proposition}[theo]{Proposition}
\newtheorem{corollary}[theo]{Corollary}
\newtheorem{algorithm}[theo]{Algorithm}
\newtheorem{protocol}[theo]{Protocol}
\newtheorem{conjecture}[theo]{Conjecture}
\newtheorem{example}[theo]{Example}
\newtheorem{ex}[theo]{Example}
\newtheorem{exercise}[theo]{Exercise}
\newtheorem{claim}[theo]{Claim}
\newtheorem{postulate}[theo]{Postulate}
\newtheorem{notation}[theo]{Notation}
\newtheorem{construction}[theo]{Construction}
\newtheorem{simulator}[theo]{Simulator}
\newtheorem{remark}[theo]{Remark}
\newtheorem{observation}[theo]{Observation}
\newtheorem{question}[theo]{Question}
\newtheorem{hypothesis}[theo]{Hipothesis}
\newtheorem{definition}{Definition}[section]
\newtheorem{theorem}{Theorem}[section]
\newtheorem{cryptosystem}[theo]{Cryptographic system}
\newtheorem{problem}[theo]{Problem}
\newtheorem{doubts}{DOUBTS}

\maketitle

\begin{center}
\noindent\textsuperscript{1} Instituto de Telecomunica\c{c}\~{o}es,\\Av. Rovisco Pais 1049-001, Lisboa, Portugal\\
\textsuperscript{2} Departamento de Matem\' atica, \\Instituto Superior T\' ecnico, Universidade de Lisboa, \\Av. Rovisco Pais 1049-001, Lisboa, Portugal\\
\textsuperscript{3}  Departamento de Inform\'atica, \\Faculdade de Ci\^encias, Universidade de Lisboa, \\ Campo Grande, 1749-016 Lisboa, Portugal\\
\textsuperscript{4}  LaSIGE, Faculdade de Ci\^encias, Universidade de Lisboa, Portugal\\
Campo Grande, 1749-016 Lisboa, Portugal
\end{center}

\begin{abstract}
We present a bit-string quantum oblivious transfer protocol based on single-qubit rotations. 
%
%The proposed protocol does not violate the Lo's no-go theorem that prevents the unconditional security of $1$-out-of-$2$ oblivious transfer.
%
Our protocol is built upon a previously proposed quantum public-key protocol and its practical security relies on the laws of Quantum Mechanics. 
Practical security is reflected in the fact that, due to technological limitations, the receiver (Bob) of the transferred bit-string is restricted to performing only ``few-qubit'' coherent measurements. 
We also present a single-bit oblivious transfer based on the proposed bit-string protocol.
The protocol can be implemented with current technology based on optics.
\end{abstract}

\newpage
\section{Introduction}
\label{sec:introduction}

Since the success of quantum cryptography, which allowed for the exchange of secret keys~\cite{ben:bra:84,eke:91,ben:92}, and whose security is based on the laws of physics, instead of the unproven mathematical conjectures, a hope appeared for designing various quantum protocols with improved security, with respect to their classical counterparts. One of the basic protocols used in building complex multiparty security schemes is the {\em Oblivious Transfer (OT) Protocol} (often referred to as all-or-nothing OT).

OT can be seen as a game played by two parties, Alice and Bob. 
Alice has many secrets that wishes to share with Bob in such a way that at the end, on average, Bob learns half of those secrets and Alice does not know which secrets Bob really knows. 
Each instance of this protocol, used to reveal in half of the cases Alice's secret, is the Oblivious Transfer Protocol.

OT consists of two distinct phases: 
	(i) the transferring phase, during which Alice sends an encoded secret information to Bob; 
	(ii) the opening phase, during which Alice reveals enough information so that Bob can decode the secret with probability $1/2$. Note that Bob knows if he got the message or not.

OT is said to be secure if the following properties hold:
	(i) the protocol is {\em concealing}, i.e., before the opening phase, Bob is not able to learn the message sent by Alice, while after the opening phase Bob learns the message with probability $1/2$;
	(ii) the protocol is {\em oblivious}, i.e., after the opening phase, Alice remains oblivious to whether or not Bob got the message.

Rabin was the first to formally present an oblivious transfer protocol in 1981~\cite{rab:81}. The security of Rabin's OT relies on the fact that factoring large integers is not known to be possible to perform in polynomial time on classical computers. 
Later, Even, Goldreich and Lempel presented a variation of this scheme called  {\em $1$-out-of-$2$ oblivious transfer}~\cite{eve:gol:lem:85}. 
The difference to Rabin's OT is that Alice sends two messages and Bob gets only one of the two with equal probability (again, Alice does not know which message Bob decoded). 
Although differently defined, Cr\'epeau showed that when the messages are single bits the two flavours of oblivious transfer protocols are equivalent, in the sense that one can be built out of the other and vice versa~\cite{cre:87}. Furthermore, one can build an $1$-out-of-$2$ oblivious transfer protocol that transmits bit-string messages from $1$-out-of-$2$ oblivious transfer protocol for single bits~\cite{bra:cre:rob:86,cre:san:93,bra:cre:san:96}.

The oblivious transfer is a building block of more complex security protocols~\cite{bra:cre:rob:86,kil:88,har:lin:93} using Yao's garbled circuits \cite{yao:86}, and various secure multiparty computation schemes~\cite{cra:dam:mau:00,lin:pin:12,lin:zar:13}.

\begin{figure}[H]
\begin{center}
\includegraphics[angle=270, width = 12cm, keepaspectratio=true] 
{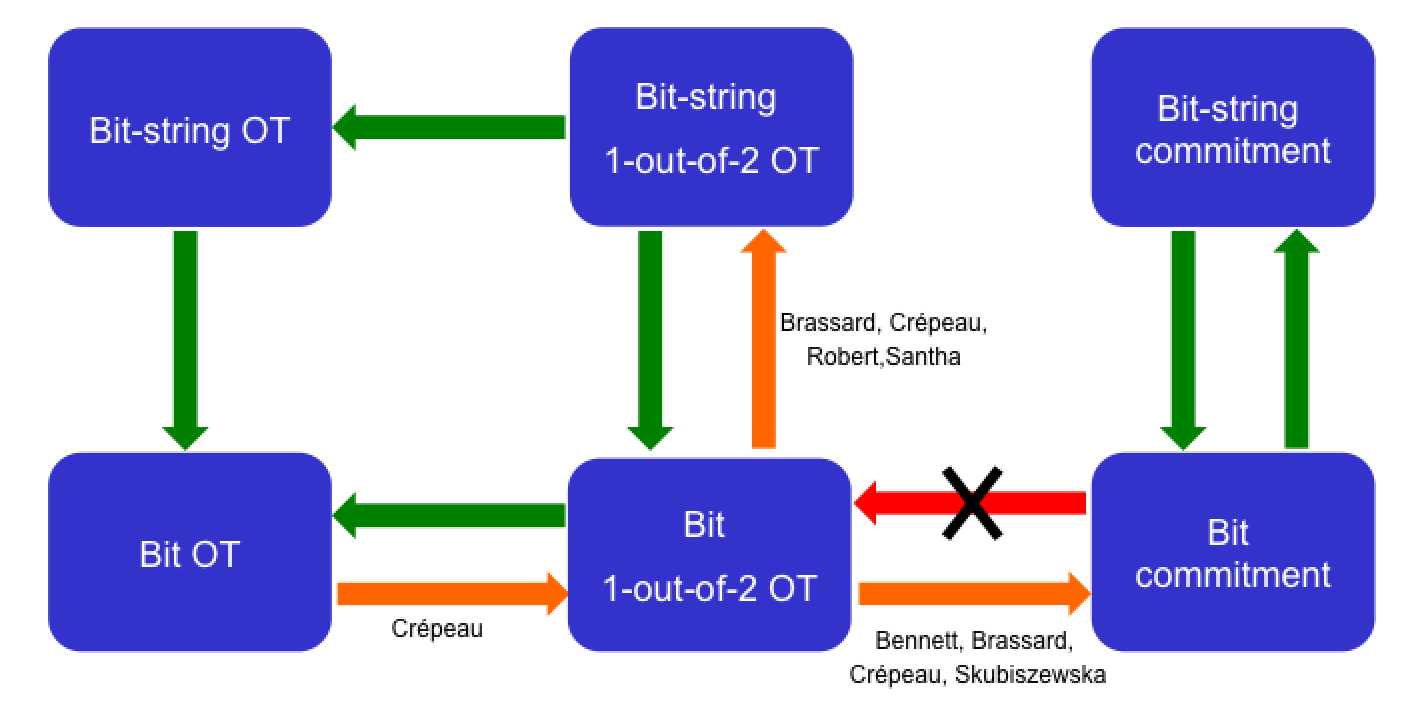}
\caption{Classical reductions between cryptographic primitives. The green arrows represent straightforward reductions; the orange ones are non-trivial reductions; the red one is the impossible implication.}
\label{fig:classicalreductions}
\end{center}
\end{figure}

Another cryptographic primitive used in designing more complex secure protocols is bit commitment~\cite{bra:cha:cre:88}. Although it is not possible to construct an OT protocol out of a bit commitment~\cite{sal:98} it was shown that  bit commitment can be reduced to $1$-out-of-$2$ bit oblivious transfer protocol~\cite{ben:bra:cre:sku:91}. In Figure~\ref{fig:classicalreductions} we schematically present the classical reductions between the above discussed cryptographic primitives.

%
%\newpage

Due to the advance of quantum computation and quantum information, the development of cryptographic applications resilient to quantum adversaries has been extensively studied in the last decades. Wiesner launched the field of quantum cryptography in $1969$  by presenting notions such as quantum money and quantum multiplexing (and only managed to publish his results in 1983~\cite{wie:83}), the latter being essentially a quantum counterpart of a $1$-out-of-$2$ oblivious transfer protocol. 

Further developing Wiesner ideas, Bennett and Brassard presented the well-known BB84 quantum key distribution protocol~\cite{ben:bra:84}, which was subsequently showed to be unconditionally secure~\cite{Lo:cha:99,sho:pre:00,may:01,sca:bsc:csr:dus:lut:pee:09}, while its classical counterparts are only computationally secure. By ``unconditionally secure'' it is meant ``secure, provided the agents have unlimited power allowed by the laws of physics''. Nevertheless, inevitable imperfections in implementations lead to various practical levels of security. In the current study, we analyse the theoretical security of our protocol, while leaving the details of particular implementations for future work. Another example of a  protocol whose quantum realisation outperforms its classical counterpart is the recently proposed contract signing protocol~\cite{pau:bou:mat:11}. 

Despite these positive results, a number of no-go theorems imposed limits to quantum cryptography. Independently, Lo and Chau \cite{lo:chau:97}, and Mayers~\cite{may:97}, showed that unconditionally secure quantum bit commitment protocol is impossible, within the scope of non-relativistic physics (for unconditionally secure protocols that use relativistic effects, see~\cite{ken:99,ken:05,bou:feh:gon:sch:13}). Subsequently, Lo~\cite{lo:97} proved similar no-go theorem for all ``one-sided two-party computations'' protocols. An immediate consequence of this result is the impossibility of having secure $1$-out-of-$2$ oblivious transfer. The alternative, ensuring practical security of such protocols, is to consider noisy or bounded memories~\cite{weh:sch:ter:08,sch:ter:weh:11,koe:weh:wul:12,ng:jos:min:kur:weh:12,bou:feh:gon:sch:13,lou:14,alm:etal:14,alm:etal:15}. 
Recently, a (quantum) computationally secure version of oblivious transfer protocol was presented in~\cite{sou:mat:ada:pau:14}.  

\begin{figure}[H]
\begin{center}
\includegraphics[angle=0, width = 14cm, keepaspectratio=true] 
{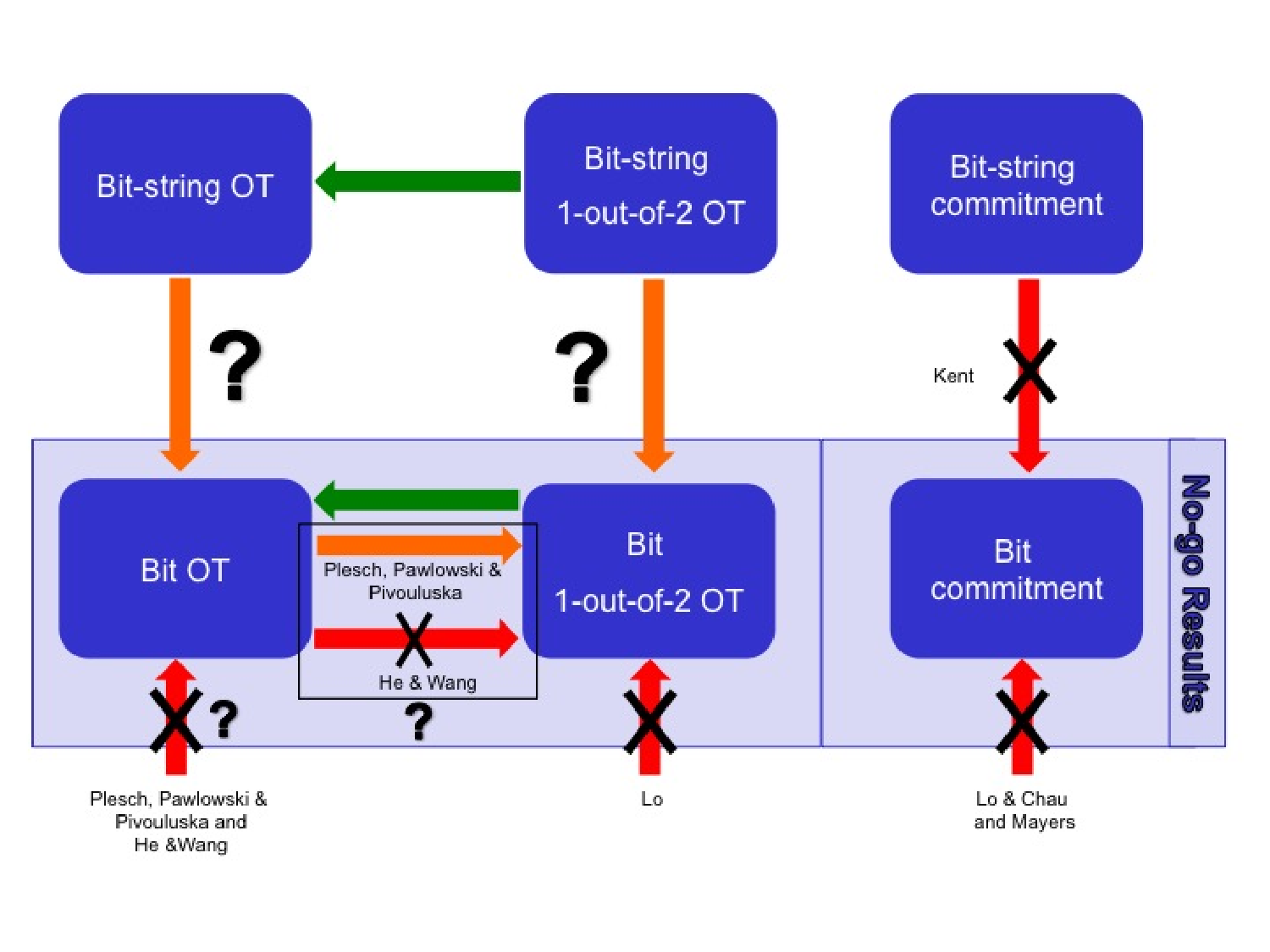}
\caption{Quantum reductions between cryptographic primitives and no-go Theorems. The green arrows represent straightforward reductions; the orange ones are non-trivial reductions; the red ones are the impossible implications. Open questions are represented with a question marks.}
\label{fig:quantumreductions}
\end{center}
\end{figure}

Following the classical equivalence~\cite{cre:87} between the two flavours of oblivious transfer, one might conclude that the impossibility of having unconditionally secure $1$-out-of-$2$ oblivious transfer would imply the same for oblivious transfer. 
But the rules of quantum physics present a wider range of possibilities, thus potentially compromising classical reduction schemes. 
Namely, as to build the $1$-out-of-$2$ oblivious transfer one has to run several oblivious transfer protocols as black boxes, the possibility of the so-called coherent attacks -- joint quantum measurements on several black boxes -- arises. 
Thus, having secure quantum oblivious transfer protocol does not necessarily mean that it is possible to construct secure $1$-out-of-$2$ oblivious transfer. 
Indeed, He and Wang recently showed that in quantum domain the various types of oblivious transfer are no longer equivalent~\cite{he:wan:06a} and constructed a secure quantum single-bit oblivious transfer~\cite{he:wan:06b} using entanglement. 
Consequently, classical reductions of single-bit to a bit-string protocols are also compromised in the quantum setting and need to be re-examined. Recent example of constructing a secure quantum bit-string commitment protocol~\cite{ken:03:b}, despite the above mentioned no-go theorems for single-bit commitment~\cite{lo:chau:97, may:97}  is yet another example of invalidity of classical reductions (see also a quantum bit-string generation protocol~\cite{bar:mas:04}). Therefore, a need of explicitly constructing quantum bit-string oblivious transfer protocol which is not based on classical reductions mentioned above arises~\cite{cre:87,bra:cre:rob:86,cre:san:93,bra:cre:san:96}. In Figure~\ref{fig:quantumreductions} we present quantum reductions between cryptographic primitives. Note that, despite the mentioned controversial (and not widely accepted within the community) results by He and Wang~\cite{he:wan:06a,he:wan:06b}, the recent construction of a single-bit $1$-out-of-$2$ oblivious transfer out of an ordinary  (i.e., all-or-nothing) OT~\cite{ple:paw:piv:16} leaves the question of the equivalence between the two flavours of the protocol open.

In this paper, we present a quantum oblivious transfer protocol for bit-strings, based on the recently proposed public key cryptosystem~\cite{nik:08}. Each bit of the string to be transferred is encoded in a quantum state of a qubit, in such a way that states corresponding to bit values $0$ and $1$ form an orthonormal basis. The key point of the protocol is that for each qubit, the encoding basis is chosen at random, from some discrete set of bases. 

In the OT protocol, the two agents do not trust each other, and so no authentication is required at this level. OT is a protocol that implements a channel to be used as a building block of more complex multi-party computations. Nevertheless, at this higher level, classical authentication might be required, which can be achieved by exchanging signed messages (using pre-shared keys) via OT (see for example~\cite{gol:02}, Section 2.3.2.3, p. 48). Alternatively, OT can be used to achieve the very authentication between the agents~\cite{can:12}. In~\ref{sec:aut}, we present details of the authentication mechanism performed at the level of multi-party computation.

The paper is organised as follows. In the next Section, we present our OT protocol, while in Section~\ref{sec:security} we analyse the protocol's security. Finally, in Section~\ref{sec:conclusions} we summarise the results and present possible future lines of research. In~\ref{sec:appendix} we present the technical details (notation, definitions and results) used in the paper.

\section{The protocol}
\label{sec:ot}

In this section, we present the protocol that achieves oblivious transfer of a bit-string message from Alice to Bob. 
The scheme uses hash functions which allow to certify if after the opening phase Bob got the message or not. 
A hash function produces a {\it digest} of a message -- a string of smaller size -- such that: (i) the probability of generating at random strings with the same hash value is negligible; (ii) the hash values are almost uniformly distributed over the set of all possible digests.

\begin{figure}[H]
\begin{center}
\includegraphics[angle=0, width = 12cm, keepaspectratio=true] 
{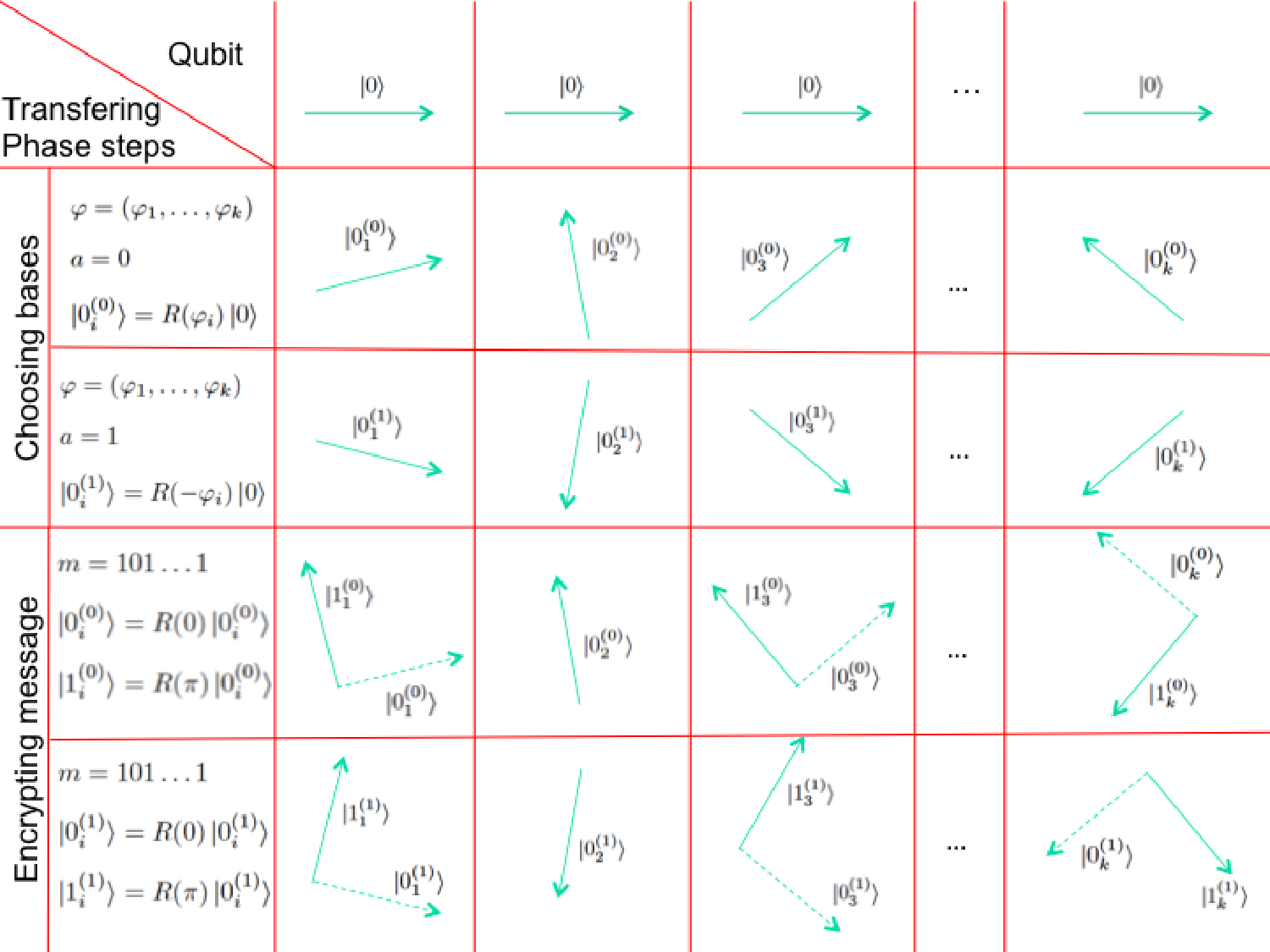}
\caption{Schematic description of the transferring phase of our oblivious transfer protocol for messages of length $k$. The full arrows represent the actual states of qubits, while the dashed arrows in the last two lines (encryption of a message) represent $\ket{0_i}$ states.}
\label{fig:transferingphaseOT}
\end{center}
\end{figure}

Our protocol is based on the public-key cryptosystem~\cite{nik:08}, and can be briefly summarised as follows (for more details, see~\ref{sec:appendix}).
Given a reference, so-called computational basis $\beta_0 = \{ \ket 0, \ket 1\}$, Alice first encodes each bit $m_i$ of the message (a string of bits) $\mathbf{m} = m_1\ldots m_k$ into the state $\ket{m_i}$ of the corresponding qubit. Then, she randomly chooses a bit value $a$, and for each $m_i$ a rotation angle $\varphi_i$ (taken from a given set of angles $\Phi$), and rotates $\ket{m_i}$ by $(-1)^a\varphi_i$. Finalising the transferring phase, she sends the qubits to Bob. Note that for each qubit $i$ the encoding quantum states 
\begin{eqnarray}
\ket{0^{(a)}_i}  & = &  R((-1)^a\varphi_i)\ket 0 \\[3mm]
\ket{1^{(a)}_i}  & = &  R((-1)^a\varphi_i)\ket 1 = R(\pi)\ket{0^{(a)}_i},
\end{eqnarray}
where rotations $R(\varphi)$ are defined by $R(\varphi)\ket 0 = \cos(\varphi/2)\ket 0 + i \sin(\varphi/2)\ket 1$, are mutually orthogonal and hence fully distinguishable, provided one knows the direction $a$ and the angle $\varphi_i$ of the rotation. Therefore, Bob cannot decipher the message $\mathbf{m}$, unless given additional information about the encoding bases $\beta_i=\{\ket{0^{(a)}_i},\ket{1^{(a)}_i}\}$. 
In Figure \ref{fig:transferingphaseOT}, we present a schematic description when the length of the message to be transferred is $k$.

%\newpage
%

%\newpage

In the opening phase, Alice provides Bob with such (partial) information: she sends the so-called secret key, a string $\varphi = (\varphi_1,\ldots, \varphi_k)$ of rotation angles, but {\em not} the rotation direction $a$. Oblivious to the rotation direction, Bob can only guess it, which he will get correctly in $50\%$ of the cases. 

Encrypted in quantum states of qubits, Alice sends the message $\mathbf{m}$, together with its digest $\mathbf{d}=h(\mathbf{m})$, given by a suitably chosen hash function $h:\{0,1\}^k\to \{0,1\}^{\omega}$, with $\omega= \lfloor \sqrt k \rfloor$. Upon decrypting the states of qubits sent by Alice, Bob recovers a string which is a concatenation of the form $\mathbf{m}'\mathbf{d}'$. Note that $\mathbf{m}'$ and $\mathbf{d}'$ are not necessarily the message $\mathbf{m}$ and its hash value  $\mathbf{d}=h(\mathbf{m})$. Bob checks if $\mathbf{d}'= h(\mathbf{m}')$. If so, he is convinced  that the received message $\mathbf{m}'$ is indeed the intended message $\mathbf{m}$ (for technical details, see Section~\ref{sec:security}). 

Below, we present a rigorous description of our bit-string OT protocol. The parameters that determine the protocol's security levels are $k,n \in \mathbbm N$, with $k$ being the length of the message, while $n$ defines the ``elementary rotation angle'' $\theta_n = \pi/2^{n-1}$.   The rotation angles are then given by $\varphi_i = s_i\theta_n$, where $s_i\in\{0,\dots, 2^{n}-1\}$. For the reasons of simplicity, we will use the term ``secret key'' for both the string of rotation angles ${\varphi} = (\varphi_1,\ldots, \varphi_{k+\omega})$, as well as the string of numbers $\mathbf{s} = (s_1,\dots,s_{k+\omega})$ that determine them. Before the opening phase, all the information about the protocol parameters are private to Alice, while the only public information is the hash function $h$ (note that knowing the hash function, one knows $k$ and $\omega$ as well). At the beginning of the opening phase, Alice (publicly) announces to Bob $n$ and $\mathbf s$. 

%The only information that also other parties have is the  kind of hash function to be used. 

%\subsection*{Bit-string oblivious transfer}

%\newpage
\begin{protocol}[Bit-string OT]\label{prot:ot}\

\begin{quote} 
\begin{description}
\item[Security parameters:] $k, n \in \mathbbm{N}$, with $\theta_n=\pi/2^{n-1}$;
\item[Message to transfer:] $\mathbf{m}=m_1\dots m_k$;
\item[Hash function:] $h:\{0,1\}^k\to \{0,1\}^{\omega}$, with $\omega= \lfloor \sqrt k \rfloor$;
\item[Secret key:] $\mathbf{s} = (s_1,\dots,s_{k+\omega})$, where each $s_i\in\{0,\dots, 2^{n}-1\}$.
\end{description}

\begin{description} 
\item[Transferring phase:] \
\begin{enumerate}
	\item Alice chooses uniformly at random the hash function $h$ and a bit $a \in \{0,1\}$  and prepares the following state (with $\mathbf{d} = h(\mathbf{m})$, and $d_i$ being its $i^{th}$ bit): 
%	\begin{eqnarray}
%	\hspace*{-45mm}
%	\ket{\psi}&
%	\hspace*{-40mm}
%	=&
%	\hspace*{-35mm} 
%		\ds
%		\bigotimes_{i=1}^k R(m_i\pi + (-1)^a\times s_i\theta_n)\ket{0} \bigotimes_{i=1}^{\omega} R(h_i(\mathbf{m})\pi + (-1)^a\times s_{i+k}\theta_n)\ket{0} \\[3mm]
%		% 
%		&\hspace*{-40mm}
%		=&
%		%
%		\ds
%		\hspace*{-35mm}
%		\left(\bigotimes_{i=1}^k \left[ \cos\left(\frac{m_i\pi+(-1)^a\times s_i\theta_n}{2}\right)\ket{0} + \sin\left(\frac{m_i\pi+(-1)^a\times s_i\theta_n}{2}\right)\ket{1}\right] \right)\otimes\\
%		%
%		&&
%		\hspace*{-35mm}
%		\nonumber\ds \left(\bigotimes_{i=1}^{\omega} \left[\cos\left(\frac{h_i(\mathbf{m})\pi+(-1)^a\times s_{i+k}\theta_n}{2}\right)\ket{0} + \sin\left(\frac{h_i(\mathbf{m})\pi+(-1)^a\times s_{i+k}\theta_n}{2} \right)\ket{1} \right]  \right)
%	\end{eqnarray}
%
%	(Note that $h_i(\mathbf{m})$ represents the $i^{th}$ bit of the binary string $\mathbf{h}(\mathbf{m})$).
	% 
\begin{eqnarray}
	{\hspace*{-125mm}
	\ket{\psi} }&
%	\hspace*{-60mm}
	=&
%	\hspace*{-15mm} 
		\ds
		\bigotimes_{i=1}^k R(m_i\pi + (-1)^a\times s_i\theta_n)\ket{0} \bigotimes_{i=1}^{\omega} R(d_i)\pi + (-1)^a\times s_{i+k}\theta_n)\ket{0} \;\;\;\;\;\;\;\;\;\; \\[3mm]
		&\hspace*{-75mm}
		=&
		\ds
		\hspace*{-40mm}
		\left(\bigotimes_{i=1}^k \left[ \cos\left(\frac{m_i\pi+(-1)^a\times s_i\theta_n}{2}\right)\ket{0} + \sin\left(\frac{m_i\pi+(-1)^a\times s_i\theta_n}{2}\right)\ket{1}\right] \right)\otimes\\
		&&
		\hspace*{-35mm}
		\nonumber\ds \left(\bigotimes_{i=1}^{\omega} \left[\cos\left(\frac{d_i\pi+(-1)^a\times s_{i+k}\theta_n}{2}\right)\ket{0} + \sin\left(\frac{h_i\pi+(-1)^a\times s_{i+k}\theta_n}{2} \right)\ket{1} \right]  \right)
	\end{eqnarray}
%
%	(Note that $\mathbf{d}_i$ represents the $i^{th}$ bit of the binary string $\mathbf{m}$).
%

	\item Alice sends the state $\ket\psi$ to Bob.
\end{enumerate}

\vspace{15mm}
\item[Opening phase:] \
	\begin{enumerate}
	\setcounter{enumi}{2}
	\item Alice sends to Bob the secret key $\mathbf{s} = (s_1,\dots,s_{k+\omega})$,  the security parameter $n$.
	\item \label{step:checking} Bob checks if $\mathbf s$ is likely to be a possible output of a random process.
	
	\item Bob chooses uniformly at random $a'\in\{0,1\}$ and applies $R((-1)^{a'}s_i\theta_n)$ to each qubit of $\ket\psi$.
	
	\item Bob applies the measurement operator 
	$M^{\otimes (k+\omega)} = (0 \times \ket 0 \bra 0 + 1\times \ket 1\bra 1)^{\otimes (k+\omega)}$.% = \sum_{\mathbf m = 0}^{(2^k-1)}\sum_{\mathbf h = 0}^{(2^\omega-1)} (\mathbf m \mathbf h) \times $.
	
	\item Let $\mathbf{m}' \mathbf{d}'$ be the message that Bob recovers (notice that here $\mathbf{d}'$ is a bit-string, a potential value of the hash, and not a function itself). He checks if $\mathbf{d}' = h(\mathbf{m}')$. If that is the case then Bob is almost sure that $\mathbf{m}' = \mathbf{m}$, otherwise he knows that $\mathbf{m}'$ is not the correct message.
	\end{enumerate}
\end{description}
\end{quote}
\end{protocol}

Notice that knowing $h(\mathbf{m})$ can potentially reveal the whole set $A_\mathbf{m}$ of the strings mapped to the same value of hash. Knowing $A_\mathbf{m}$ decreases Bob's uncertainty about the unknown string $\mathbf{m}$, thus effectively revealing $\omega = \lfloor \sqrt k\rfloor$ bits of information about string $\mathbf{m}$. This information may help Bob to increase the probability of finding $\mathbf{m}$, thus compromising the security of the protocol. 
Therefore we encrypt both the message $\mathbf{m}$ and $h(\mathbf{m})$ into a quantum state sent by Alice. Since, in order to confirm that he obtained the message $\mathbf{m}$, Bob needs to learn the value $h(\mathbf{m})$ as well, one can consider the concatenated string $\mathbf{m}h(\mathbf{m})$ as a message to be transferred. 
For simplicity, in the rest of the paper we will denote $\mathbf{m}h(\mathbf{m})$ as a single message $\mathbf{m}$ to be transferred. Note though, that there are correlations between the message $\mathbf{m}$ and the value $h(\mathbf{m})$, which might become relevant for the Concealing property, and in particular for achieving the Probabilistic transfer, after Bob learns the particular function $\mathbf{m}$ chosen by Alice. We will address this issue when discussing the above mentioned cases.

In Step $4$ Bob checks if the secret key $s$ was indeed randomly chosen.  By encoding $s_i$'s into binary numbers Alice has to provide an $n\times (k+\omega)$ long bit-string produced by a fair coin. A number of possible tests of random-number generators exist in literature, such as $\chi^2$, Kolmogorov-Smirnov, Serial correlation, Two-level, K-distributivity, Serial and Spectral tests (for more details, see~\cite{jai:91}, Chapter~$27$).  
Step 4 of the protocol is used to overcome the hypothetical chance of Alice to cheat by sending particular elements $s_i$ of the secret key $s$ which allow Bob to recover the message with probability close to $1$. 
Notice that if $s_i$ is such that the angle of rotation is $\varphi_i = s_i\theta_n/2 = 0$, or $\varphi_i = s_i\theta_n/2 = \pi/2$, then Bob will with certainty get the correct bit value $m_i$. 
Therefore, if the elements $s_i$ of the secret key were close to $0$ or $\pi/2$, Alice would know with probability significantly higher than $1/2$ that Bob received the message $\mathbf{m}$ (for a detailed analysis of possible cheating strategies of Alice, see below the proof of the obliviousness criterion). 
If all $s_i$'s were indeed chosen uniformly at random, than significant portion of them would not be close to $0$ nor $\pi/2$, preventing Alice from cheating. 
%So, Bob checks the degree of randomness of the secret key $s$ produced by Alice: he wants to be sure that the majority of the elements $s_i$ will not allow for (almost) certain message transfer. 

Nevertheless, for the protocol to be secure, a much simpler criterion can be used, one that is satisfied whenever a string is indeed produced uniformly at random. If Alice chooses each $s_i$ uniformly at random, then on average half of such choices satisfy $\varphi_i = s_i\theta_n/2 \in [\pi/8,3\pi/8] \cup [5\pi/8,7\pi/8]$. These $s_i$'s are already far enough from $0$ and $\pi/2$ to secure the protocol against cheating Alice. For a detailed discussion on the degree of Bob's confidence against cheating  strategies of Alice, see below the proof of the obliviousness criterion.

%implying a negligible overhead on the Bob recovering the message and the knowledge of Alice about the fact that an honest Bob got or not the message. These statistical tests provide an easy way for  Bob evaluate the degree of confident that Alice is playing fairly and the advantage she has regarding the acknowledgment of the message by him, and decide accordingly to abort the protocol or not. We provide a detailed discussing on the degree of confidence of Bob and the respective probability of Alice in due course, when we analyze the obliviousness of the protocol.

Typically, bit-string protocols achieve required security levels for {\em suitably large} lengths of messages, as is the case in Kent's protocol~\cite{ken:03:b}, for example (for more details see Section~\ref{sec:security}). Nevertheless, having a secure bit-string OT protocol by encrypting single bits by sufficiently large messages, one can design secure single-bit OT. Below, we present a simple way of using our protocol to achieve oblivious transfer of a single bit $b$ by sending a bit-string message $\mathbf{m}$. 

%\clearpage
\begin{protocol}[Single-bit oblivious transfer]\label{prot:1/2-OT}
\  
\begin{quote}
\begin{description}
\item[Message to transfer:] $b$;
\item[Security parameters:] $k$ and $n$;
\end{description}

\begin{enumerate}
	\item Alice chooses bit $b$.
	\item Alice chooses a $2k$-bit message $\mathbf{m}$, such that $\ds \bigoplus_{i=1}^k m_i \times m_{k+i} = b$.
	\item Alice and Bob perform Protocol \ref{prot:ot}.
	\item If Bob had got the right message $\mathbf{m}$, then he performs $\ds\bigoplus_{i=1}^k m_i \times m_{k+i} = b$. Otherwise, he cannot recover the bit.
\end{enumerate}

\end{quote}
\end{protocol}

\section{Security analysis}
\label{sec:security}

In this section, we analyse the security of our oblivious transfer protocol. 
%The protocol's security depends on parameter $n$ through Theorem $5$, which is satisfied asymptotically, i.e. for {\em sufficiently large} values of $n$. The dependence on $\ell$ comes through Probabilistic transfer requirement (see below Point $2$ of Theorems $14$ and $15$), stated in terms of negligible functions $\varepsilon(\ell)$, again for {\em sufficiently large} values of $\ell$.~\footnote{A function $\varepsilon(\ell)$ is said to be negligible if  $\varepsilon(\ell)<1/p(\ell)$ for any polynomial $p(\ell)$ and sufficiently large~$\ell$.} Note that, since we require  that $n\geq\ell$, the security of the protocol is stated in terms of a negligible function  in $\ell$.~\footnote{In fact, it is enough to have $n = O(\ell)$. To simplify the presentation, we take $n\geq\ell$.} 
Oblivious transfer has to satisfy the following four properties (the first express the correctness while the last three assure the security of the protocol):

\begin{description}

\item[\hspace{10pt} Soundness:] If both Alice and Bob are honest, then with probability $1/2$ Bob will obtain the right message. While Bob knows if he got the right message or not, Alice is oblivious of that fact.

\item[\hspace{10pt} Concealingness:] If Alice is honest Bob cannot learn the whole message that Alice meant to send before the opening phase (the protocol is concealing). 

\item[\hspace{10pt} Probabilistic transfer:] After the opening phase, Bob cannot learn the whole message in more than $50\%$ of the cases (with probability higher than $1/2$).

\item[\hspace{10pt} Obliviousness:] If Bob is honest then Alice does not know if Bob received the message -- she can only guess with probability $1/2$  (the protocol is oblivious).
\end{description}

In case of bit-string protocols, the probability $1/2$ that appears in the above definition of Soundness and Probabilistic transfer properties (but not Obliviousness) is relaxed to ${1}/{2} + \varepsilon(k)$, where $k$ is the length of the message and $\varepsilon:\mathbbm{N} \rightarrow \mathbbm{R}$ is a negligible function, i.e., for every positive polynomial $p$ there exists a $k_0 \in \mathbbm{N}$ such that for all $k > k_0$, $\varepsilon(k) \leq {1}/{p(k)}$ (for the definition and a detailed overview of the use of the notion of negligible functions in cryptography, see for example \cite{gol:04}).

Note that Bob can get {\em a part} of the message $\mathbf{m}$ with certainty, as long as the number of additional bits that he has to guess is larger than a polylogarithmic function of $k$ (a function is polylogarithmic if it can be written in the form $f(x) = \sum_i c_i\log^{p_i} (x)$, for fixed constants $c_i$ and powers $p_i$, see for example~\cite{bla:04}). This means that the uncertainty of the entire message is greater than any polylogarithmic function of $k$. For instance, in the case a cheating Bob can learn all but $\sqrt{k}$ bits of the message, the probability of him guessing properly the {\em whole} message is $2^{-\sqrt{k}}$ (i.e., Bob's uncertainty of the whole message is $\sqrt{k}$). This way, Alice can use our protocol to send $\sqrt{k}$ bits of information of a shorter string $\mathbf{i}$ whenever the transfer is successfully achieved; otherwise, Bob's probability to learn the intended information $\mathbf{i}$ is negligible (e.g., of the order of $2^{-\sqrt{k}}$; note that the quadratic increases in the resources is common and acceptable in implementation of communication protocols). To do this, she can use an error-correcting code that can correct up to $\sqrt{k}$ errors. Such code consists of cells  (``codewords'') each of size $2^{\sqrt{k}}$. By randomly choosing the cell, and encoding $\mathbf{i}$ in one of its $2^{\sqrt{k}}$ elements, Bob is left without virtually any knowledge of $\mathbf{i}$ whenever the OT is unsuccessful: by being able to learn $(k-\sqrt{k})$ bits of $\mathbf{m}$ he can at most identify the cell, but not $\mathbf{i}$ (the uncertainty of the received string $\mathbf{m}$ is precisely $\sqrt{k}$). Note the importance of stating the security criterion in terms of negligible functions, i.e., in terms of asymptotic behaviour, with respect to the message length (be it $k$ of $\mathbf{m}$ or, through the error-correcting code, $\sqrt{k}$ of $\mathbf{i}$), of the probabilities to learn (parts of) the information sent, which is a standard approach in cryptography and computer science (see for example~\cite{gol:04}, where negligible functions are introduced already in the Introduction).

In general, both quantum and classical cryptographic security protocols for exchanging messages depend on several parameters, one of them being the length of the message. As a rule, such protocols are said to be secure if, provided that the other parameters are suitably chosen, the cheating probability is negligible with respect to the length of the message, which is satisfied asymptotically, i.e., for {\em sufficiently large} values of $k$.

In our case, as well as in the case of the public-key scheme presented in~\cite{nik:08} (on which our protocol is based; see also the related recent scheme based on quantum walks~\cite{vla:rod:mat:pau:sou:15}), one such parameter is $n$, and for both protocols the level of security indeed depends on the choice of $n$. Nevertheless, as proven in~\cite{nik:08}, with a proper choice of $n$, the public-key scheme is secure against eavesdropping. Consequently, with the same choice of a proper $n$, our protocol is Sound (correct), Concealing (before the opening phase Bob cannot learn the message sent by Alice) and achieves Probabilistic transfer (on average, Bob receives half of the messages sent by Alice). On the other hand, the last  security criterion (Obliviousness) does not depend on the choice of $n$, as shown in the respective proofs presented below.

Note that in \cite{nik:08}, in order to further reduce the probability of a successful attack, security parameter $n$ was treated as a part of the secret key (together with $s$). But it was noted that the protocol would still be secure even if $n$ were public. In a subsequent paper \cite{sey:nik:alb:12}, in which the robustness of the public-key cryptosystem introduced in \cite{nik:08} was further analysed, $n$ was treated as a part of a public key, i.e., the cryptosystem is secure even if (a properly chosen) $n$ were known. Note that in both cases, according to the above definition, the protocol is secure, but with different negligible functions $\varepsilon(k)$: when $n$ is private, the corresponding negligible function is smaller than when $n$ is public.

Definitions of Soundness, Concealingness and Probabilistic transfer properties have somewhat weak requirement of Bob not being able to learn the {\em whole} message $\mathbf{m}$ before and in (about) half of the cases after the opening phase. According to this definition, even if Bob were able to learn all but, say the last bit of the message, the protocol would still be secure. One can adopt a stronger criterion, requiring that only with negligible probability Bob can learn {\em with certainty} part of $\mathbf{m}$ which would allow him to {\em infer}, with {\em non-negligible probability}, the rest of the message. In other words, to require that the acquired information cannot help Bob to {\em improve} his chances of guessing $\mathbf{m}$, in a sense that before and after receiving quantum (and classical) key(s), Bob's chances to correctly guess the whole $\mathbf{m}$ are negligible in the length of the message. Such security criterion is in the spirit of the one presented in the classic paper~\cite{eve:gol:lem:85} introducing 1-out-of-2 oblivious transfer protocol.

Note that even in the case of the correct inference of the rest of the message, a cheating Bob can still not know {\em with certainty} whether the guess was correct, which is weaker position than that of an honest player who knows with certainty if the message is received. Note that by sheer guessing, Bob will always, on average, correctly infer half of the bits of $\mathbf{m}$, but without knowing which bits he guessed right. To say that Bob knows certain part of the message means to know both the values and the corresponding positions of the known bits. Thus, there is a difference between knowing the message (as required by the definition of the protocol), and being right in inferring (guessing) the message. 

As we show below, regarding Concealingness, our protocol satisfies the strong\-est possible requirement: no part of the message can be known at all (with certainty). Indeed, upon obtaining a string of qubits from Alice, Bob can {\em at best} learn the whole quantum state of the system received. Since each quantum state $\ket \psi$ sent by an honest Alice can, for a suitable choice of the secret key $\mathbf{s} = (s_1,\dots,s_{k+\omega})$, encode {\em any} string $\mathbf{m}\mathbf{d}\in \Sigma^{k+\omega}$, before the opening phase Bob is completely clueless of any part of the message sent. 

Regarding Probabilistic transfer, we show that, providing that Bob is constrained to perform only ``few-qubit'' coherent measurements, he can at best with  probability of $1/2$ learn all but the first $\sqrt{k}$ bits of the message. Nevertheless, the probability to successfully infer the missing $\sqrt{k}$ bits is $2^{-\sqrt{k}}$, i.e., negligible in the length of the message. By ``few-qubit'' coherent measurements we mean that Bob cannot perform a joint  measurement on more than a given finite number of qubits, a reasonable practical  constraint for today's and any not-so-far future technology.

\subsection{Soundness of the protocol}
\label{sec:soundness}
In the following we prove the soundness of our protocol: 
%\begin{theorem}\label{the:soundness}
if both parties are honest, 
then with probability $1/2+\varepsilon(k)$ 
Bob will get the right message, 
where $\varepsilon(k)$ is negligible function on the size of the message $\mathbf{m}=m_1\dots m_k$. 
%\end{theorem}

%\begin{proof}
First assume that Alice and Bob had chosen to rotate the state in opposite directions, i.e., $a\neq a'$. Without loss of generality assume that Alice chooses $a=0$, to rotate clockwise all the qubits. The qubits Alice sent to Bob are in the following state:
\begin{eqnarray}
	\ket{\psi} 	&=& \ds
		\bigotimes_{i=1}^{k} R(m_i\pi + s_i\theta_n)\ket{0} \\[3mm]
		 	&=& \ds
		\bigotimes_{i=1}^{k}\cos\left(\frac{m_i\pi+s_i\theta_n}{2}\right)\ket{0} + \sin\left(\frac{m_i\pi+s_i\theta_n}{2}\right)\ket{1}.
\end{eqnarray}
In the opening phase Bob receives from Alice the additional information, the secret key $s=(s_1,\ldots, s_{k})$.

By the assumption, Bob decides to rotate each qubit received from Alice counterclockwise ($a'=1$) by $-s_i\theta_n$. The states he gets are either $\ket{0}$ or $\ket{1}$. In fact:
\begin{eqnarray}
	\ds R(-s_i\theta_n)( R(m_i\pi + s_i\theta_n)\ket{0})
		&=& R(m_i\pi )\ket{0}\\[3mm]
			&=&\ds\cos\left(\frac{m_i\pi}{2}\right)\ket{0} + \sin\left(\frac{m_i\pi}{2}\right)\ket{1}\\[3mm]
			&=&\ket{m_i}.
	\end{eqnarray}
Bob measures $M = 0 \times \ket 0 \bra 0 + 1\times \ket 1\bra 1$ on the above state and the result is $m_i$ with probability $1$ (note that for $m_i= 0$ we have $\cos(m_i\pi /2) = 1$, and analogously for $m_i = 1$).
We conclude that if Bob chooses to rotate in the direction contrarily to Alice's choice, then with probability $1$ Bob will recover the bit sent by Alice. 
	
On the other hand, if Alice and Bob decide to rotate each qubit of the message in the same direction ($a=a'$), say clockwise, the qubits' states are transformed into ($i=1\dots  k$):

\begin{eqnarray}
	\hspace{-10mm}\ds R(s_i\theta_n)( R(m_i\pi \!+ \!s_i\theta_n)\ket{0})&=& 
			\ds R(m_i\pi + 2s_i\theta_n)\!\ket{0})\\[3mm]
			&=&\ds\cos\left(\!\frac{2 s_i\theta_n+m_i\pi}{2}\!\right)\!\ket{0}\! + \!\sin\left(\!\frac{2s_i\theta_n+m_i\pi}{2}\!\right)\!\ket{1}\\
			&=& \ket{\tilde m_i}.
	\end{eqnarray}
If $m_i=0$ then the  above state becomes 
$
\ket{\tilde m_i}=\ds\cos\left(s_i\theta_n\right)\ket{0} + \sin\left(s_i\theta_n\right)\ket{1}
$ 
and by measuring the qubit with $M = 0 \times \ket 0 \bra 0 + 1\times \ket 1\bra 1$
Bob gets the correct answer with probability 
$\cos^2(s_i\theta_n)$; 
if $m_i=1$ then the above state becomes 
$
\ket{\tilde m_i}=
	-\sin\left(s_i\theta_n\right)\ket{0} + 
	\cos\left(s_i\theta_n\right)\ket{1}
$
and again Bob gets the correct bit with probability 
$\cos^2(s_i\theta_n)$. 
Hence
\begin{eqnarray}
\ds \Pr\left(m_i; M, \ket{\tilde m_i}\right) 
		%=& \ds\frac 1 2 \cos^2(s_i\theta_n) + \frac 1 2  \cos^2(s_i\theta_n)\\ 
		= \ds \cos^2(s_i\theta_n).
\end{eqnarray}

Assuming that the key $\mathbf{s}$ is chosen at random, the probability of recovering the whole message by rotating in the wrong direction becomes negligible, and the expected probability of recovering message $\mathbf{m}$, when measuring $M^{\otimes k}$, on the state 
$
\ket{\psi^\prime} = \bigotimes_{i=1}^{k} R((-1)^{a'}s_i\theta_n) \ket{\psi}
$
is:
\begin{eqnarray}
\Pr(\mathbf{m}; M^{\otimes k},\ket{\psi^\prime}) 
		\!\!&=&\!\!\Pr( a'\!\not =\! a) \!\times\! \Pr(\mathbf{m}| a' \!\not=\! a) \!+\! 
			\Pr( a' \!= \!a) \!\times\! \Pr(\mathbf{m}|a'\!=\!a)\\[3mm]
		&\leq& \ds \frac 1 2 + \frac 1 2 \prod_{i=1}^k\cos^2(s_i\theta_n).\label{uppr}
\end{eqnarray}

%Clearly,  when Alice chooses the values $s_i$ at random, the expected probability of Bob recovering the message $m$ in case Alice and Bob perform equal rotations becomes negligible, i.e., $\varepsilon(k)= \ds\frac 1 2 \prod_{i=1}^k\cos^2(s_i\theta_n)$  is negligible.
%
%To see this, notice that on average half of values for the rotation angles $s_i\theta_n/2$ fall in the region $[\pi/8;3\pi/8] \cup [5\pi/8;7\pi/8]$, giving the upper bound $\varepsilon(k)\leq 2^{-k/2}$. 

The two cases, $a' \not = a $ and $a'=a$, occur both with probability $1/2$. While in the first case Bob always gets $\mathbf m$ correctly, in the second, the probability given by the random values $s_i$ chosen by Alice, is given by $\ds\prod_{i=1}^k\cos^2(s_i\theta_n)$. To see that $\varepsilon(k)= \ds\frac 1 2 
   \prod_{i=1}^k\cos^2(s_i\theta_n)$ is negligible, notice that on average half of values for the rotation angles $s_i\theta_n/2$ fall in the region $[\pi/8;3\pi/8] \cup [5\pi/8;7\pi/8]$,  for which $\cos(s_i \theta_n)<\sqrt{2+\sqrt 2}/2$, giving the upper bound $\varepsilon(k)\leq 2^{-k/2}$. 

\begin{figure}[H]
\begin{center}
\includegraphics[angle=0, height = 5cm, keepaspectratio=true] 
{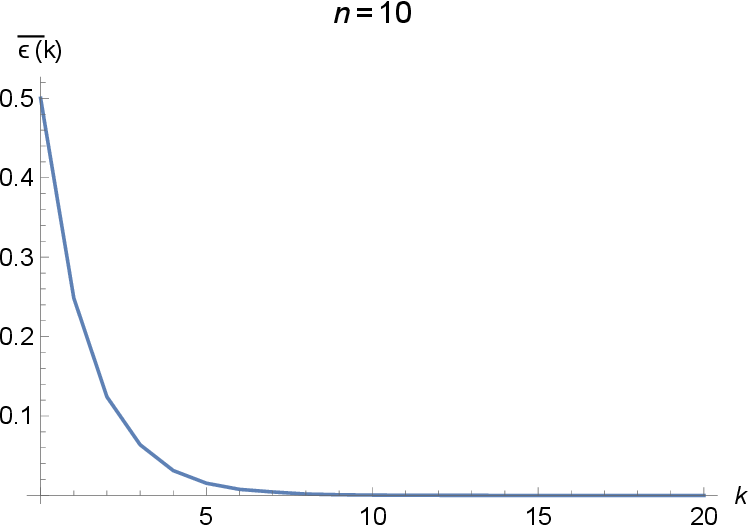}
\caption{The expected value $\bar{\varepsilon}(k)$, for $n=10$ and $100$ randomly chosen strings~$\mathbf s$.}
\label{fig:n10prob}
\end{center}
\end{figure}

Given $\theta_n$, one can estimate the average value $\bar{\varepsilon}(k)$ as a function of the message length $k$, thus obtaining the security level of the protocol's soundness criterion. The current technology available at the market allows for the accuracy of the state of polarisation (SOP) of about $0.2^\circ$ on Poincar\'e sphere (see the accuracy specifications of the ``Polarization Instrumentation'' at~\cite{thorlabs:17}). This roughly corresponds to the case of $n=10$, for which on Figure~\ref{fig:n10prob} we plot the expectation value $\bar{\varepsilon}(k)$, for $100$ randomly chosen strings $\mathbf s$. As one can see, already  for the messages of four letters the ``access'' probability $\bar{\varepsilon}(k)/2$ is of the order of 0.01. 

The information received by Bob consists of two parts: one corresponding to the actual message sent by Alice, and the other corresponding to its hash value. At the end of the protocol, Bob checks if he recovered the correct message by comparing its hash value with the latter part of information received. Note that by the properties of hash functions, the probability that the hash of the first part matches the second one is negligible in the case Alice and Bob performed the same rotation (see~\ref{sec:appendix} for more detailed description of the properties of hash functions).
%\end{proof}

\subsection{Concealingness of the protocol}
\label{sec:concealingness}
%We proceed by proving the concealing property of the protocol. 

In this subsection we show that 
%
%\begin{theorem}
if Alice is honest, the probability of Bob recovering Alice's message before the opening phase is negligible. Furthermore, after the opening phase Bob recovers the message with, up to a negligible value, probability $1/2$.
%\end{theorem}

%\begin{proof}

The first part of the statement follows directly from the security of the public-key cryptosystem~\cite{nik:08} (see the discussion on one-way functions and state distinguishability in~\ref{sec:appendix}), and is basically a consequence of the fact that, depending on the secret key component $s_i$, the same state $\ket{\psi_i}$ of a single qubit can be encrypting either a $0$ or a $1$: for each $s_i$ there exists $s_i'$ such that $\ket{\psi_i} = R(s_i\theta_n) \ket 0$ encrypts $0$, while $\ket{\psi_i} =  R(s_i'\theta_n+\pi)\ket 0$ encrypts $1$. In fact, before the opening phase, our protocol is as secure as the cryptosystem underlying it.

We stress that the additional information provided by Alice, the hash function cannot help Bob recovering the message $\mathbf{m}$. In fact, below, we prove that even if Bob had access directly to the hash value $\mathbf{h}(\mathbf{m})$ this would not help him (note that since the value $\mathbf{h}(\mathbf{m})$ is encrypted makes Bob's task even harder). 
In the following, we provide the reasoning for a particular hash function.

Given a message $\mathbf{m}$, consider its partition into $\omega = \lfloor\sqrt k\rfloor$ consecutive blocks of bits $\mathbf{\bar{m}}_i$ ($i = 1, \ldots \omega$), each with length $\lfloor\sqrt k\rfloor$: $\mathbf{m} = \mathbf{\bar{m}}_1 \ldots \mathbf{\bar{m}}_\omega$. 
Each bit $d_i = h_i(\mathbf{m})$ of the hash value $\mathbf{d} = h(\mathbf{m})$ is the parity of the $i$-th block of the message $\mathbf{m}$: $d_1 = h_1(\mathbf{m}) = m_{1}\oplus \ldots \oplus m_\omega$, etc. Hence, all the bits of $\mathbf{d} = h(\mathbf{m})$ are mutually independent. 

Suppose that $\mathbf{d} = h(\mathbf{m})$ allows to recover $\mathbf{m}$ with some non-negligible probability $p$. Then, in particular, the bit $d_1 = h_1(\mathbf{m})$ helps to recover the possible block $\mathbf{\bar{m}}_1 = m_{1}\oplus \ldots \oplus m_{\omega}$, with the same probability $p$. We claim that this is impossible, assuming that the cryptosystem \cite{nik:08}, used to design our protocol, is secure. 

In fact, if a cryptosystem is secure for coding a message $\mathbf{m}$ of length $k$, then {\em a fortiori} the encryption of a polynomially shorter message, say $\mathbf{\bar{m}}_1$, is also secure. So, if $d_1 = h_1(\mathbf{m})$ would help to recover the first block with non negligible probability $p$ then, by randomly guessing the value $d_1 = h_1(\mathbf{m})$ (that will be correct with probability $1/2$), it would be possible to break the cryptosystem presented in \cite{nik:08} with non-negligible probability $p/2$. 

One can easily describe other hash functions  by considering all possible forms of dividing  $k$ elements into groups of $\omega$ elements, i.e., by using the above hash function $h$ on the permuted message. Given a permutation $\pi \in S_k$ of length $k$, one can define the hash function $h_\pi (\mathbf{m}) = h_{id_k} (m_{\pi(1)} \dots m_{\pi(k)})$, where $h_{id_k}$ is the above $h$. Obviously, the concealing property is valid for all hash functions in the set $\{ h_\pi | \pi \in S_k\}$.

\subsection{Probabilistic transfer of the protocol}
\label{sec:prob_transf}

After receiving the secret key $\mathbf{s}$, Bob's description of the qubits sent by Alice is given by the mixed state (for convenience, we consider $a \in \{+,-\}$, where ``$+$'' stands for clockwise rotation and ``$-$'' otherwise):
\begin{eqnarray}
%\hspace{-25mm}
\rho_B(\mathbf{s}) = \frac 1 2 \!\!\!\sum_{a \in \{+,-\}} \!\!\!\left(\frac 1 2\right)^k \!\!\!\!\!\!\sum_{m_1 \in \{0,1\}} \!\!\!\ldots\!\!\! \sum_{m_k \in \{0,1\}} \ket{m_1(s_1)}_a\!\bra{m_1(s_1)}\otimes\ldots\otimes\ket{m_{k}(s_{k})}_a\!\bra{m_{k}(s_k)},
\end{eqnarray}
where $\ket{m_i(s_i)}_{\pm} = \cos\left(\frac{m_i\pi}{2} \pm \frac{s_i\theta_n}{2}\right)\ket{0} +  \sin\left(\frac{m_i\pi}{2} \pm \frac{s_i\theta_n}{2}\right)\ket{1}$.
The single-qubit partial states are completely mixed, and can be written in the following suitable form: $\rho_B(s_i) = \frac 1 2 (\rho_0(s_i) +\rho_1(s_i))$, where $\rho_{m_i}(s_i) = \frac 1 2 (\ket{m_i(s_i)}_+\!\bra{m_i(s_i)} +\ket{m_i(s_i)}_-\!\bra{m_i(s_i)})$. 
Note though that the overall state $\rho_B(\mathbf{s})$ is not a tensor product of single-qubit states: the rotation direction $a$ is the same for all qubit thus correlating single-qubits. Nevertheless, if Bob is constrained to perform only few-qubit coherent measurements, these correlations, as well as the knowledge of $\mathbf{d} = h(\mathbf{m})$, cannot help him to increase the probability of learning $\mathbf{m}$.

First, we give the proof for the case of single-qubit measurements. As before, the hash function $h$ is determined by the parity of blocks $\mathbf{\bar{m}}_i$ of size $\omega$. Since the parity of block $\mathbf{\bar{m}}_i$ is completely uncorrelated to the value of each of its bits, unless we know the values of all other $\omega - 1$ remaining bits, the choice of the optimal single-qubit measurement of at least $\omega - 1$ qubits of a single block does not depend on the hash value $d_i = h_i(\mathbf{m})$.

The correlations between single-qubit states established by the same choice of the rotation direction cannot help either. A possible cheating strategy would be to, as prescribed by the protocol, randomly choose the rotation direction, and perform the corresponding measurement on first few qubits only. With probability $1/2$ the choice will be right, and the bits would be correctly decrypted; with probability $1/2$ though, the wrong choice would lead to wrong decryption which, in case Bob can detect it, would result in measuring the right observable on the remaining qubits. But Bob can detect the wrong choice only by comparing the results with the hash value, the parity of blocks of length $\omega$. Thus, only upon measuring all qubits of at least one block of size $\omega$ Bob can spot the mistake. This however leaves him uncertain which, among $2^{\omega - 1}$ possible messages, was the message sent by Alice, which are exponentially many on the size $k$ of the whole message $\mathbf{m}$ (note that $\omega = \sqrt{k}$). Thus, since for each $s_i$ the states $\rho_0(s_i)$ and $\rho_1(s_i)$ are not fully distinguishable, what Bob can do is to try to distinguish between the two states as best as possible. 

\begin{figure}[H]
\begin{center}
\includegraphics[angle=0, height = 5cm, keepaspectratio=true] 
{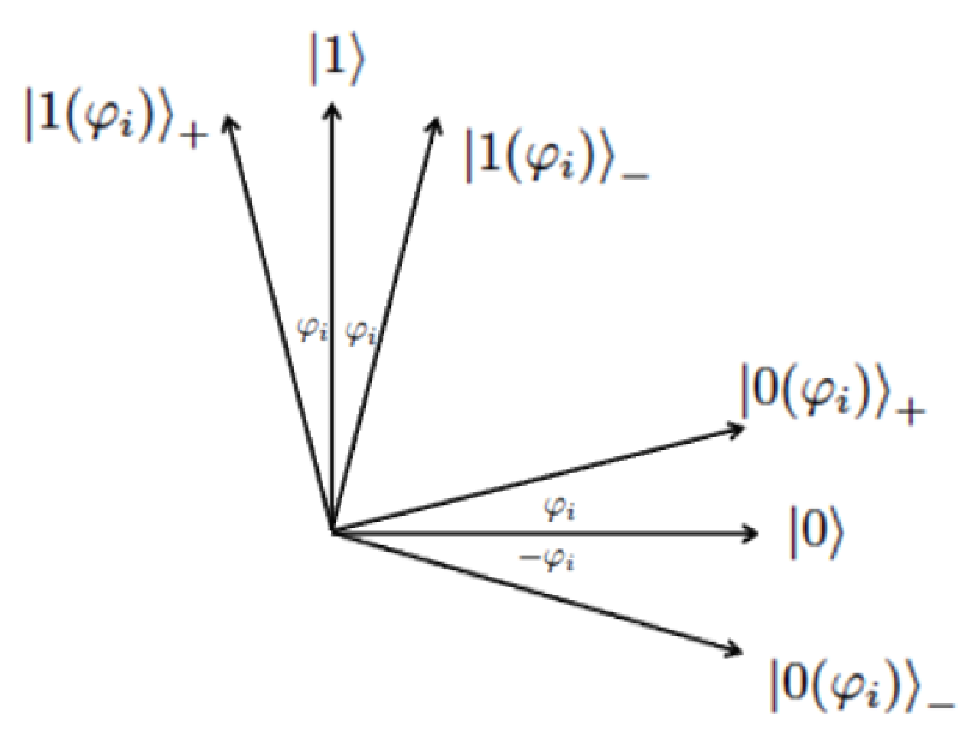}
\caption{The optimal discrimination between the bit values $0$, encoded in quantum state $\rho_{0}(s_i) = \frac 1 2 (\ket{0(s_i)}_+\!\bra{0(s_i)} +\ket{0(s_i)}_-\!\bra{0(s_i)})$, and the bit value $1$, $\rho_{1}(s_i) = \frac 1 2 (\ket{1(s_i)}_+\!\bra{1(s_i)} +\ket{1(s_i)}_-\!\bra{1(s_i)})$. The optimal observable is given by the vectors from the computational basis, $\ket 0$ for inferring the bit value $0$, and  $\ket 1$ for inferring the bit value $1$. Note that $\varphi_i = s_i \theta_n /2$.}\label{fig:states}
\end{center}
\end{figure}

The optimal probability of guessing bit's value $m_i$ is then given by the Helstrom formula~\cite{hel:69}: 
\begin{eqnarray}
\label{eq:cheating_probabilistic}
\mbox{P}_{H}(\rho_0(s_i), \rho_1(s_i)) = \frac 1 2 + \frac 1 4 \mbox{Tr} |\rho_0(s_i)- \rho_1(s_i)| = \frac 1 2(1 + |\cos(s_i\theta_n)|).
\end{eqnarray}
Note that the optimal observable for such measurement is the same for each possible $s_i$, and is given by the computational basis $\{ \ket 0 , \ket 1 \}$ (see Figure \ref{fig:states}).
Analogously as in the proof of soundness of the protocol, since on average half of values $s_i$ satisfy $|\cos(s_i\theta_n)| \leq 1/\sqrt{2}$, we have $\varepsilon (k) \leq q^{-k/2}$, where $q = \frac 1 2 (1 + {1} / {\sqrt{2}}) < 1$. It follows that the probability to guess any non-negligible number of bits is still negligible (by ``negligible number of bits'' we mean the number of bits such that, knowing their values the probability to guess the rest of $\mathbf{m}$ is negligible).

%\newpage
%

%
%\newpage

%
%Therefore, the only way for Bob to recover the message $m$ is to follow the protocol and choose direction $a'$ at random, in which case he obtains $m$ with probability $1/2$. 
%
%Note that Bob cannot guess rotation direction $a$ with probability bigger than $1/2$, as the mixed states corresponding to either direction are completely indistinguishable. Indeed a single-qubit state can be written as  $\rho_B(s_i) = \frac 1 2 (\rho_+(s_i) +\rho_-(s_i))$, where 
%
%$\rho_\pm(s_i) =\frac 1 2 (\ket{0(s_i)}_\pm\!\bra{0(s_i)} +\ket{1(s_i)}_\pm\!\bra{1(s_i)}) = \I /2$.
% \end{proof}

%The above proof is valid for single qubit measurement of a cheating Bob. We conjecture that the protocol is secure against multi-qubit measurements as well. Indeed, if Bob were able to, using coherent multi-qubit measurements, learn the message sent by Alice, then for sufficiently large $n$ and $k$ he would be able to distinguish between virtually any two quantum states.

% (for more details on coherent attacks in quantum cryptography, see~\cite{may:95,big:boy:boy:mor:roy:00,sho:pre:00,bra:lut:mor:san:00,bih:boy:bra:gra:mor:02}).}

Suppose now Bob is allowed to perform at most two-qubit coherent measurements. Then, for each pair, say $(s_1,s_2)$, the four quantum states
\begin{eqnarray}
\rho_{00} (s_1s_2) &=& 
\ds\frac 1 2 (\rho^+_{00} (s_1s_2) + \rho^-_{00} (s_1s_2))\\[3mm] &=&
 \ds \frac 1 2 (\ket{0(s_1)0(s_2)}_+\!\bra{0(s_1)0(s_2)} + \ket{0(s_1)0(s_2)}_-\!\bra{0(s_1)0(s_2)}),
\end{eqnarray}
(and analogously for $\rho_{01} (s_1s_2)$, $\rho_{10} (s_1s_2)$ and $\rho_{11} (s_1s_2)$), would also not be fully distinguishable. Therefore, the optimal strategy that Bob can adopt will produce wrong decryption, with finite error probability $q>0$. As in the case of single-qubit measurements, this leads to negligible advantage over the $1/2$ probability of recovering $m$, given sufficiently large $k$ (and thus the block length $\omega = \sqrt{k}$).

Given the maximal length $\ell$ of the multi-qubit measurement, each block $\tilde{m}$ of length $\ell$
%, that can be measured coherently, 
is from Bob's point of view described by the mixed state $\rho_{\mathbf{\tilde{m}}}(\mathbf{\tilde{s}}) = \frac 1 2 (\rho^+_{\mathbf{\tilde{m}}}(\mathbf{\tilde{s}}) + \rho^-_{\mathbf{\tilde{m}}}(\mathbf{\tilde{s}})) = \frac 1 2 (\ket{\mathbf{\tilde{m}}(\mathbf{\tilde{s}})}_+\!\bra{\mathbf{\tilde{m}}(\mathbf{\tilde{s}})} + \ket{\mathbf{\tilde{m}}(\mathbf{\tilde{s}})}_-\!\bra{\mathbf{\tilde{m}}(\mathbf{\tilde{s}})})$, where $\mathbf{\tilde{s}}$ is the part of the secret key $\mathbf{s}$ corresponding to the block $\mathbf{\tilde{m}}$. 
%The range of $\rho_{\tilde{m}}(\tilde{s})$ is the two-dimensional Hilbert space spanned by the states $ \ket{\tilde{m}(\tilde{s})}_+$ and $\ket{\tilde{m}(\tilde{s})}_-$ that are used to encode the block $\tilde{m}$: $\mathcal{H}_{\tilde{m}}(\tilde{s}) = \mbox{span} \{ \ket{\tilde{m}(\tilde{s})}_+ , \ket{\tilde{m}(\tilde{s})}_- \}$. 
As $\ell$ increases, the states $\ket{\mathbf{\tilde{m}}(\mathbf{\tilde{s}})}_\pm$ and $\ket{\mathbf{\tilde{m}}'(\mathbf{\tilde{s}})}_\pm$, corresponding to two different messages $\mathbf{\tilde{m}}$ and $\mathbf{\tilde{m}}'$, become increasingly distinguishable. The precise relation between the maximal length $\ell$ of the allowed coherent measurements and the size $k$ of the message $m$ will be addressed in a separate study. 

To summarise, the crucial nontrivial part of the proof of the probabilistic transfer criterion restricted to few-qubit measurements is to show that the probability to learn $m$ is negligible, i.e., it scales faster than the inverse of any polynomial on the length of the message. Since the few-qubit states of the ``maximal length'' $\ell$ on which Bob performs the measurements are not fully distinguishable, he will infer correctly that part of the message only with probability $p_\ell$ strictly  smaller than $1$. Thus, upon repeating the finite-length few-qubit measurements $t=k/\ell$ times, the probability of correctly inferring the whole message scales exponentially to zero as $p_{\ell}^t = p_\ell^{k/\ell} = (\sqrt[\ell]{p_\ell})^k$, which, given the behaviour of $p_\ell$ (a subject of a future study) provides the explicit relation between the message length $k$ and the maximal measurement length $\ell$, for a desired security level. Thus, given the message length $k$ and the desired security level $\varepsilon$ (upper value for the probability to cheat), one can determine the fixed ``maximal length'' $\ell$ of allowed multi-qubit joint measurement by solving the equation $\varepsilon = (\sqrt[\ell]{p_\ell})^k$ on $\ell$. Note that by ``multi-qubit joint measurement'' we mean any kind of generalised POVM allowed by the laws of quantum mechanics. Further, to perform the analysis of the relation between $k$ and $\ell$, it is enough to find a suitable upper bound of $p_\ell < 1$.

\subsection{Obliviousness of the protocol}
\label{sec:oblivious}
To finish the security discussion we prove that the protocol is oblivious: 
%.  In the sequel we prove  
%
%\begin{theorem}
%The Protocol \ref{prot:ot} is oblivious, 
%{\it i.e}.,  
at the end of the protocol Alice does not know whether Bob received the right message of not.
%\end{theorem}

%\begin{proof}
%At the end of the protocol, since Bob performs local operations and measurements, Alice has no way of knowing if Bob had chosen the right rotation, or not. Therefore, if being honest and sending the state prescribed by the Protocol, Alice cannot know if an honest Bob received the message or not. 

At the end of the protocol, since Bob performs {\em only} local operations and measurements, {\em without communicating} to Alice, she has no way of knowing if Bob had chosen the right rotation, or not. Note that no entanglement {\em without} subsequent communication can be used to exchange/acquire information. Otherwise, one could achieve faster-than-light communication, thus explicitly violating causality and the principle of relativity. Indeed, what entanglement affects are the correlations: Bell inequalities are given in terms of various correlation functions, and the violation of local realism can be observed only upon distant observers exchange the results of their local measurements. The prominent bit-commitment cheating strategy discovered independently by Lo and Chao~\cite{lo:chau:97} and Mayers~\cite{may:95} does involve entanglement, but in addition to performing a suitable measurement on her halves of entangled pairs, a cheating Alice has to, {\em subsequently}, send her measurement outcomes to Bob. Therefore, if being honest and sending the state prescribed by the Protocol, Alice cannot know if an honest Bob received the message or not.

To increase her probability of knowing if Bob received the message or not, a cheating Alice can try sending a cheating state $\ket{\psi_{ch}}$ that would, with probability significantly higher than $1/2$ (ideally, with probability $1$), reveal $\mathbf{m}$ independently of Bob's choice of rotation (obviously, in order to maintain the $50\%$ of Bob's success, Alice has to send a statistical mixture of different cheating states, some having the mentioned ``success probability'' significantly higher than $1/2$, the others significantly lower). Note that in our protocol {\em no entanglement} between Alice and qubits sent to Bob could possibly help her to cheat, by adapting the so-called Mayers-Lo-Chau type of attack, which indeed enables Alice to successfully cheat during the execution of a quantum bit-commitment protocol (the essence of the mentioned no-go theorem~\cite{lo:chau:97,may:97}). In order to evade the {\em binding criterion} of the bit-commitment protocol (the requirement that after making the commitment to a bit value, it is not possible to change it during the opening phase of the protocol), Alice establishes perfect correlations between her and Bob's results for {\em one of the two different choices} of local measurement bases (computational for committing to $0$, and diagonal for committing to $1$), i.e., during the opening phase she informs Bob which measurement to perform. In our protocol, if using entanglement to cheat, the choice of pairs of local observables, for which perfect (or at least, better than random) correlations between Alice's and Bob's measurement results is achieved, is irrelevant. Thus, Alice performing local measurements on entangled states is equivalent to locally preparing at her site a simple statistical mixture of the corresponding collapsed Bob's states, and then sending them to him. Thus, the entanglement strategy is indeed equivalent to sending simple statistical mixture of non-entangled states. Below, we show that for every quantum state $\ket{\psi_{ch}}$ of Bob's particle, the probability to obtain the intended message (and thus the probability of Alice's inference that Bob obtained the message) is bounded from above by a function that is exponentially close, with respect to the message length $k$, to $1/2$.

Let $l$ be the number of $s_i$'s for which $\varphi_i = s_i\theta_n /2 \in [ \pi/8; 3\pi/8 ] \cup [ 5\pi/8; 7\pi/8 ]$. For such cases we can consider the rearranged secret key $\mathbf{s}=s_1\dots s_l$ and the corresponding message $\mathbf{m}=m_1\dots m_l$. Depending on his choice of rotation direction $a'$ Bob will measure one of the two observables $C_{\pm}(\mathbf{s}) = \sum_{\mathbf{m}=0}^{2^l-1} \mathbf{m}\times P_{\pm}(\mathbf{m};\mathbf{s})$, where one-dimensional projectors are given by $P_{\pm}(\mathbf{m};\mathbf{s}) = \bigotimes_{i=0}^{l}P_{\pm}(m_i;s_i) = \bigotimes_{i=0}^{l} \ket{m_i(s_i)}_\pm\!\bra{m_i(s_i)}$ and messages (bit-strings) $\mathbf{m}$ are seen as binary numbers.

For given $\mathbf{m}$ and $\mathbf{s}$ Alice wants to maximize the probability $\Pr_{ch}$ of Bob obtaining $\mathbf{m}$ measuring $C_{\pm}(\mathbf{s}) $ on $\ket{\psi_{ch}}$ (and thus her probability of knowing if he got the message or not), which is given by
\begin{eqnarray}
\label{eq:cheating_obliviusness}
{\Pr}_{ch} = \frac 1 2 \left( ||P_{+}(\mathbf{s}) \ket{\psi_{ch}}||^2 + ||P_{-}(\mathbf{s}) \ket{\psi_{ch}}||^2 \right).
\end{eqnarray}
The state $\ket{\psi_{ch}}$ that maximizes the above expression is  the equal superposition of $\ket + = \ds\prod_{i=1}^l \ket{m_i(s_i)}_+$ and $\ket - = \ds\prod_{i=1}^l \ket{m_i(s_i)}_-$. Thus, we have:
%In order to maximize the probability of cheating Alice must choose to send a state which rotation is in the middle of the two possible outcomes and therefore, it is easy to see that:
\begin{eqnarray}
{\Pr}_{ch} \leq \frac 1 2 \left(1+|\braket{+|-}|\right) \leq \frac 1 2 \left(1 + \cos^{l} (\pi/8)\right).
\end{eqnarray}

If the values $s_i$ were produced uniformly at random, then the probability that  $\varphi_i = s_i\theta_n /2 \in [\pi/8;3\pi/8] \cup [5\pi/8;7\pi/8]$ is $1/2$. As a consequence, the random variable that counts the number $l$ of such $\varphi_i$'s follow the binomial distribution $\mathcal B (k,1/2) $, with $k$ being the number of trials (the total number of rotation angles $\varphi_i$, equal to the length of the message $\mathbf{m}$) and $1/2$ being the success probability of each trial (where by ``success'' we mean that the rotation angle falls within the above mentioned intervals). For sufficiently large $k$, the binomial distribution can be approximated by the normal distribution $\mathcal N(\mu,\sigma^2)$ with the mean $\mu= k/2$ and the variance $\sigma^2= k/4$. This allows Bob to set the degree of confidence of Alice's obliviousness. For example, choosing the $3\sigma$ criterion, if $(k-3\sqrt k)/2\leq l \leq (k+3\sqrt k)/2$  Alice's probability to learn if Bob got the message or not will be ${\Pr}_{ch} = 1/2 + \varepsilon(k)$, where $\varepsilon(k)$ is negligible (which happens in $99.8\%$ of the cases if $s_i$ were chosen uniformly at random). 
%\end{proof}
%in $\Pr[(k-3\sqrt k)/2\leq l \leq (k+3\sqrt k)/2]= 99.8\%$ of the cases

Regarding the quantitative evaluation of the security levels against the cheating agents in terms of $\theta_n$ and $k$, we note that one can perform the estimation analogous to the one performed for the soundness criterion, presented above in Section~\ref{sec:soundness}. Indeed, the security levels, given through the cheating probabilities, are for the probabilistic transfer and the (single-qubit measurements) obliviousness criteria given by~(\ref{eq:cheating_probabilistic}) and~(\ref{eq:cheating_obliviusness}), respectively, which have the similar dependence on the values $\cos(s_i \theta_n)$, for a given $\mathbf s$, to that of $\varepsilon (k)$ from~(\ref{uppr}). Regarding the concealingness criterion presented in Section~\ref{sec:concealingness}, 
%its security is based on the Holevo theorem and is a direct consequence of the public-key cryptosystem~\cite{nik:08}.
while estimating its numerical security in terms of $\theta_n$ and $k$ is, for general coherent multi-qubit measurements, a non-trivial task, one can straightforwardly see that the numerical expressions for the security levels for the case of single- and few-qubit measurements would again involve expressions similar the the ones considered above, given in terms of $\cos(s_i \theta_n)$.

Finally, we note that, according to the security criterion adopted in this paper, our OT protocol is secure against violating {\em only one} out of those three criteria (concealingness, probabilistic transfer and obliviousness), while {\em keeping the other two satisfied}. In case a cheating Alice decides not to send a message at all (by sending a wrong secret key $s$), she knows with certainty that Bob will not receive it (the protocol is not oblivious), at the same time violating the probabilistic requirement. This compromises the mentioned Cr\'epeau's reduction~\cite{cre:87} to a 1-out-of-2 OT, as the cheating Alice would know with probability 1/2 which of the two messages Bob chose to receive~\cite{he:15}. Nevertheless, using the reduction presented in \cite{cho:etal:09} and a bit-commitment protocol (such as those presented in~\cite{ken:99,ken:05,ng:jos:min:kur:weh:12,lou:14}), one can achieve a  1-out-of-2 OT through the Cr\' epeau's reduction based on a protocol such as ours (see \cite{sou:mat:ada:pau:15} for a detailed discussion on the example of the computationally secure OT presented in\cite{sou:mat:ada:pau:14}). Moreover, one can use our protocol to achieve two-party computation even without the help of a bit commitment. The semi-honest 1-out-of-2 OT based on our OT protocol is insecure against a cheating Alice. Due to the symmetry of the OT protocol~\cite{wol:wul:06}, it can be further transformed into a semi-honest 1-out-of-2 OT insecure against a cheating Bob, who would with probability 1/2  obtain both messages sent by Alice (and only one in the remaining half of the cases). This is nothing but the ``cut-and-choose OT primitive'', introduced in~\cite{lin:pin:11} and further developed in~\cite{Lin:16}, which allows for two-party computation (for cut-and-choose protocols on garbled circuits~\cite{yao:86}, see also~\cite{lin:pin:07,moh:riv:13,she:11}). 

Note that the security of the complex multi-party protocols is at most as high as the security of their building blocks: for a complex scheme to have certain level of security, its building blocks have to maintain the same security level, but this is not enough -- a complex scheme could be insecure against, say, coherent measurements performed over all of its building blocks, etc. 
%In the case of the reductions using our protocol, they are provable secure as long as only few-qubit measurements are performed. 
Our protocol is proven secure only against ``few-qubit measurements'', i.e., it is only practically secure, its security being guaranteed by the technological limitations.  Thus, our reduction to a single-bit OT, as well as the above mentioned application~\cite{lin:pin:11} to achieve multi-party computation without the use of BC, are proven to be only practically, and {\em not unconditionally}, secure. Note that the latter is a classical reduction, thus further compromising its security which, according to Lo-Chau-Mayers no-go results, cannot be unconditional.
%: for a complex scheme to have certain level of security, its building blocks have to maintain the same security level, but this is not enough -- a complex scheme could be insecure against, say, coherent measurements performed over all of its building blocks, etc. 
As a consequence, no matter how high the security level our bit-string OT has, this fact {\em alone} is not in contradiction with Lo-Chao-Mayers no-go theorems.% on the impossibility of unconditionally secure multi-party computation.

In case it turns out that our protocol is indeed insecure against general coherent attacks, such result would support the impossibility of unconditionally secure single-bit all-or-nothing OT (anticipated by the significant portion of the community), and be in tune with the proven no-go theorems, in connection to the above-mentioned multi-party computation scheme~\cite{lin:pin:11}. 
%(impossibility of unconditionally secure multi-party computation). 
Nevertheless, from the technical point of view, it would be interesting to see the particular coherent attack.

In case it turns out that our protocol is secure against general coherent attacks, it would not directly contradict no-go theorems, as our proposal is a bit-string all-or-nothing OT which, similarly to a bit-string commitment, in quantum domain does not necessarily have to be equivalent to a single-bit counterpart. Nevertheless, the question of the security level of our reduction from a bit-string to a single-bit all-or-nothing OT becomes an interesting one. In this case, two possible outcomes are: (i) similarly to the case of the application to multi-party computation~\cite{lin:pin:11}, our reduction is insecure against coherent attacks (for example, proving the security against learning the string of bits does not necessarily mean that a cheating Bob cannot learn its parity instead, thus breaking the reduction security), or (ii) our single-bit all-or-nothing OT is  secure against coherent attacks. Note that, while the predominant opinion within the community is that such security level is impossible, there exists an opposite controversial result~\cite{he:wan:06b}, accompanied by the concrete construction of allegedly secure all-or-nothing OT~\cite{he:wan:06a}. Despite the recent proposal of the reduction from all-or-nothing to 1-out-of-2 OT~\cite{ple:paw:piv:16}, there exists no broadly accepted clearcut proof of the equivalence of the two flavours of the single-bit OT.

To summarise, our result is not in contradiction to any no-go theorem, and while our bit-string OT protocol is probably insecure against coherent attacks, even the opposite is allowed by the known impossibility proofs.

%Nevertheless, having a protocol such as ours, together with a bit-commitment protocol (such as those presented in~\cite{ken:99,ken:05,ng:jos:min:kur:weh:12,lou:14}), using the reduction presented in \cite{cho:etal:09} one can achieve an all-or-nothing OT secure against a wider range of cheating strategies, such as the one in which, by never sending the intended message, a cheating Alice violates the obliviousness criterion while at the same time decreasing to zero Bob's probability to receive the message (see \cite{sou:mat:ada:pau:15} for a detailed discussion on the example of the computationally secure OT presented in \cite{sou:mat:ada:pau:14}).

\section{Conclusions}\label{sec:conclusions}

In this paper we proposed a novel scheme for obliviously transferring a bit-string message from Alice to Bob. The scheme presented does not violate the Lo's no-go theorem \cite{lo:97} and its security is based on the laws of quantum physics.

We proved that the protocol is secure against {\it any} cheating strategy of Bob before the opening phase (the protocol is concealing) as well as against a cheating Alice (it is oblivious, providing Alice maintains the probabilistic transfer).  Furthermore, we proved that it satisfies probabilistic transfer, provided  Bob performs only ``few-qubit'' measurements, bounded by a certain upper bound $\ell$ (the protocol is practically probabilistic). Although intuitively our protocol should, at least for sufficiently large $n$, be secure against multi-qubit measurements, a detailed analysis of its security against Bob's coherent attacks remains to be done (similarly as for the case of recently proposed and performed quantum signature protocols \cite{dun:wal:and:14,col:etal:14}). 

% We conjecture that, similarly to quantum key distribution schemes, our protocol is unconditionally secure against a general multi-qubit measurements.}

Our protocol does not use entanglement and its optical implementation could be performed using today's technology. Indeed, as already mentioned, the current technology can perform single-photon polarisation rotations for the values on $n=10$ (and possibly higher), which, using the standard photon emission and detection techniques (used, for example, in quantum key distribution) can achieve OT protocols with short time differences between the transferring and the opening phases. To achieve longer time differences, one would need to use stable long-term quantum memories, which are beyond the current technology. Nevertheless, even the  OT protocols with short times between the two phases have a number of useful examples of multi-party computation applications, such as private data mining, zero-knowledge proofs, etc. With the future advances of the technology, one can expect significant improvements in producing stable quantum memories which could, potentially, allow for our protocol to be used in applications that require for longer time-differences, such as quantum e-voting. 

Finally, we discuss the need for the use of hash functions. Recall that at the end of the protocol Bob must be sure if he got the intended message or not. This property is guaranteed by comparing the computed hash value of the received message $\mathbf{m}$ with the presumed hash value sent by Alice together with $\mathbf{m}$. Such acknowledgment of the validity of the message decoded by Bob could be done differently. Suppose that out of all possible messages (PM), Alice is constrained to send $\mathbf{m}$ from a smaller set of messages (VM), such that verifying that $\mathbf{m}$ is in VM can be easily done, but only Alice knows the elements of VM. 
Note that in order to keep the probability of receiving a message from Alice to $1/2$, up to a negligible term, the size of VM must be exponentially smaller than the size of PM.
For example,  VM could be the set of solutions to a hard mathematical problem, say 3-SAT problem. Alternatively, the message sent might be written in an existing human language, say English, making it easily recognisable by any English-language speaker.

Apart from (dis)proving the concealingness security criterion against the general coherent attacks (performing a measurement on all qubits at once), future lines of research include formulating other quantum security protocols that use single-qubit rotations to encode bit values into quantum states taken from a number of different bases. One such immediate application is in designing a quantum bit-string commitment protocol and compare it with the existing proposals. Furthermore, similarly  when generating (randomised) secret keys, single-qubit rotations could be used in creating undeniable signatures. 

In the proof of obliviousness of the protocol, Section~\ref{sec:oblivious}, we considered as relevant the encoded basis for which $ \phi_i = s_i \theta_n /2 \in [\pi/8,3\pi/8]\cup [5\pi/8,7\pi/8]$. This might suggest that one can modify the protocol by using only those angles as possible encryption bases. Note though that the concealingness of the protocol is based in the security of the public-key criptographic system of~\cite{nik:08}, which requires the whole range of equidistant angles. We leave this as an interesting question for future research. Another interesting possibility would be to further restrict the angles used, to those corresponding only to the computational and the diagonal bases, thus sending only BB84 states. Note though that such protocol would differ from our original idea in one important issue, that might affect its security. Namely, for the choice of the states from the computational bases, the cases of $a=0$ and $a = 1$  would be indistinguishable (indeed was one of the reasons to exclude the case $\phi \approx 0$ from consideration in the proof of obliviousness of our protocol). We leave this interesting possibility as a topic for a future research.

An important future line of research is to go beyond the ``proof of principle'' presented in this paper, and analyse quantitative effects of imperfect sources, noise and measurement errors in the protocol's optical realisations. In performing such study of the protocol's quantitative security levels in realistic applications, one can straightforwardly generalise the techniques used in the study of noise and measurement errors of a practical two- and four-state bit-commitment protocols, presented in~\cite{lou:14,lou:16}.

\section*{Acknowledgments}
The authors acknowledge the support of SQIG -- Security and Quantum Information Group, the Instituto de Telecomunica\c{c}\~oes Research Unit, ref. UID/EEA/50008/2013, the IT project QbigD funded by FCT PEst-OE/EEI/LA0008/2013,  and the FCT project Confident PTDC/EEI-CTP/4503/2014. A.S. also acknowledges the FCT Pos-doc scholarship SFRH/BPD/76231/2011 during which the major part of the work was done and LaSIGE Research Unit, ref. UID/CEC/ 00408/2013.

\appendix

%\section*{Appendix A}
\section{Authentication in multiparty computation}
\label{sec:aut}

Oblivious transfer is a building block used to construct secure multi-party computation. Loosely speaking, the secure multi-party computation of a function $g(i_1,\dots,i_k)=(o_1,\dots,o_k)$ is a protocol among $k$ agents, which do not trust each other. The aim of the protocol is to jointly compute function $g$ such that, each agent $A_\ell$, with $\ell\in \{1\dots k\}$:
\begin{itemize}
	\item inputs its private message $i_\ell$, which is kept secret to the other agents;
	\item receives the private output $o_\ell$, known only by $A_\ell$. 
\end{itemize}
An example of a secure multi-party computation is private data mining, where there are two agents ($k=2$), say Alice and Bob, and Alice wishes to perform some statistics $s$ over a private database $d$ of Bob. Thus, the function to compute is of the form $g(\bot,d)=(s(d),\bot)$, where $\bot$ means the empty string. 

If Bob requires Alice to authenticate herself, in order to perform the statistics, a simple way to enforce the authentication is for Alice to input a password $w$, and the statistics will only be given if the password is correct, that is, if the hash of the password coincides with the stored password hash $h_A = h(w)$ of Alice in Bob's server (similar to computer login),
\begin{equation}
	g(w,(d,h_A))=
	\left\{\begin{array}{ll}
		(s(d),\bot)  &\mbox{if }h(w)=h_A\\[6mm]
		(\bot,\bot)  &\mbox{otherwise.}
	\end{array}\right.
\end{equation}

This idea can be generalised to any secure multi-party computation, assuming that each party shares the hashes of the passwords of each all other agents. And so, instead of computing the function $g(i_1,\dots,i_k)=(o_1,\dots,o_k)$, one computes the function
\begin{equation}
	g((i_1,w_1,\mathbf{h}_1),\dots, (i_k,w_k,\mathbf{h}_k))=
	\left\{\begin{array}{ll}
		(o_1,\dots, o_k)  &\mbox{if }h(w_1)=h_{A_1}\wedge\dots \wedge h(w_k)=h_{A_k}\\[3mm]
		(\bot,\dots,\bot)  &\mbox{otherwise,}
	\end{array}\right.
\end{equation}
where $\mathbf{h}_\ell=(h_{A_1},\dots,h_{A_{\ell-1}},h_{A_{\ell+1}},\dots,h_{A_{k}})$ is the string of hashes of the passwords of all other agents but $A_\ell$. More details can be found in \cite{gol:02}.

\section{Notation, definitions and results}
\label{sec:appendix}

In this Appendix, we provide notation, necessary definitions and results for stating and proving the security of our proposal. First, we give the definition of {\em quantum one-way functions}, based on~\cite{lu:fen:05,nik:08}.

%\begin{definition}
A {\em quantum one-way function} is a map 
$f: N \rightarrow \mathcal{H}$, 
where 
$N \subset \mathbbm{Z}$ 
and 
$\mathcal{H}$ 
is a Hilbert space, such that:
\begin{enumerate}
\item $f$ is easy to compute: 
	there is a polynomial-time (in the number of bits of the input $s$) quantum algorithm that computes $f(s)\in \mathcal{H}$, 
	with $s\in N$;
\item Hard to invert: 
	without any additional information, inverting $f(s)$ non-negligibly on the input size is impossible by fundamental physical laws of quantum (information) theory.
\end{enumerate}
If inversion is computed in polynomial-time using some additional information, called the trapdoor, then $f$ is called a {\em trapdoor quantum one-way function}. In a sense, the trapdoor is a ``key'' to unlock the input  $s$.
There are several candidates for quantum one-way functions studied in~\cite{GC01,BCWW01} (a slightly different variation of quantum one-way function, with input being quantum as well, was considered in~\cite{lu:fen:05}). 
Recently, another candidate for a quantum one-way function was proposed in~\cite{nik:08}. This function considers qubit rotations  $R$ and is given by
\begin{eqnarray}
f(s) = 
	R( s\theta_n)\ket{0} =   
		\cos\left({s \theta_n}/{2}\right)\ket{0} +
		\sin\left({s \theta_n}/{2}\right)\ket{1}
\end{eqnarray}
where $s \in \{0, \dots, 2^n-1\}$ and $\theta_n = {\pi}/{2^{n-1}}$, for some fixed $n$, and $\{\ket 0,\ket 1\}$ is a fixed computational basis (i.e., $f$ is not a function of a quantum state). 
Notice that this is a quantum one-way function because:
\begin{itemize}
\item Qubit rotations $R(s \theta_n)$ can be easily implemented up to an arbitrary accuracy by a quantum algorithm involving an universal set of gates 
	(\cite{nie:chu:04}, \cite{nik:08}).
\item Due to Holevo bound, 
the maximal amount of information that can be extracted by means of a POVM on a single qubit is 1 bit. 
Since $s$ has $n$ bits, 
it is impossible to recover $s$ from a single qubit in the state 
$R(s \theta_n)\ket{0}$.
\end{itemize}

Moreover,  $f$ can be used to construct a quantum trapdoor one-way function $F(s,b)$, where $s$ is the trapdoor information for learning an unknown bit $b$~\cite{nik:08}:
\begin{eqnarray}
F(s, b) = R( b \pi) f(s) = R( b \pi) R(s\theta_n) \ket{0} = R(s\theta_n + b \pi) \ket{0}.
\end{eqnarray}
Note that inverting $F$ (learning both $s$ and $b$) is at least as hard as inverting $f$. 
Also, the ensemble of qubits, each in a state $F(s_i, b_i)$, where $s_i$ and $b_i$ are random, is described by a complete mixture 
$\rho = \I/2$, 
if $s_i$ and $b_i$ are unknown~\cite{nik:08}. 
Therefore, every binary measurement that could be used to infer unknown bit $b$ would give completely random value. Nevertheless, if $s$ is known, by applying the rotation $R( -s \theta_n)$ to $F(s, b)$ and measuring the result in the computational basis, one obtains $b$ with certainty. Therefore, $F(s,b)$ is a polynomial quantum trapdoor one-way function.

Based on the above discussion, we present the secure public-key cryptosystem proposed in~\cite{nik:08}:

\begin{protocol}[Public-Key Encryption Scheme]\label{prot:PK}
\  
\begin{quote}
\begin{description}
\item[Message to transfer:] $\mathbf{m}=m_1\dots m_l$ with $l\leq k$;
\item[Security parameter:] $n$;
\item[Secret key:] $\mathbf{s} = (s_1,\dots,s_k)$, where each $s_i\in\{0,\dots, 2^{n}-1\}$;
\end{description}

\begin{description} 
\item[Public Key Generation:] \
\begin{enumerate}
	\item For all $1 \leq i \leq k$, Alice chooses uniformly at random $s_i~\in~\{0,\dots, 2^{n}-1\}$, and $\mathbf{s} = (s_1,\dots,s_k)$ will be her private key.
	\item Alice generates the corresponding public key:
	\begin{eqnarray}
	\ket{\psi} 	&=& \ds
		\bigotimes_{i=1}^k R( s_i\theta_n)\ket{0}  \\[3mm]
		 	&=& \ds
		\bigotimes_{i=1}^k \left( \cos\left(\frac{s_i\theta_n}{2}\right)\ket{0} + \sin\left(\frac{s_i\theta_n}{2}\right)\ket{1}\right).
	\end{eqnarray}
\end{enumerate}

\vspace{3mm}
\item[Encryption:] \
	\begin{enumerate}
	\setcounter{enumi}{2}
	\item Bob wishes to send message $\mathbf{m} = m_1, \dots m_l$ where $l \leq k$.
	\item Bob obtains Alice's public key, $\ket{\psi}$.
	\item Bob encrypts his message $\mathbf{m}$ (padded with $0$ if necessary) as follows
	\begin{eqnarray}
	\hspace*{-25mm}\ket{\psi(\mathbf{m})} &=& \ds
	\bigotimes_{i=1}^k R( m_i \pi) \ket{\psi}  \\[3mm]
	 	\hspace*{-25mm}&=& \ds
	\bigotimes_{i=1}^k \left( \cos\left(\!\!\frac{s_i\theta_n}{2} \!+\! \frac{m_i \pi}{2}\!\!\right)\ket{0} + \sin\left(\!\!\frac{s_i\theta_n}{2} \!+\! \frac{m_i \pi}{2} \!\!\right)\ket{1}\right).
	\end{eqnarray}
	
	\item Bob sends $\ket{\psi(\mathbf{m})}$ to Alice.
	\end{enumerate}

	\vspace{5mm}
	
	\item[Decryption:] \
	\begin{enumerate}
	\setcounter{enumi}{6}
	\item Alice uses private key as follows
	\begin{eqnarray}
	\hspace*{-15mm}\ket{\psi'(\mathbf{m})} &=& \ds 
	\bigotimes_{i=1}^k R( -s_i\theta_n) \ket{\psi(\mathbf{m})}\\[3mm] 
	&=& \ds  
	\bigotimes_{i=1}^k \left( \cos\left(\frac{m_i \pi}{2}\right)\ket{0} + \sin\left( \frac{m_i \pi}{2} \right)\ket{1}\right)\\[3mm]
	&=& \ds  
	\bigotimes_{i=1}^k \ket{m_i}.
	\end{eqnarray}
	
	\item Alice performs measurements on each $\ket{m_i}$ in the computational basis.
	\end{enumerate}
	
\end{description}
\end{quote}

\end{protocol}
Obviously, the Public-Key Generation corresponds to the computation of $f(s)$, the Encryption phase computes $F(s,b)$ and the Decryption phase corresponds to the inversion of $F(s,b)$ with the trapdoor information $s$, which allows to learn message $\mathbf{m}$. 
As discussed in Sections IV.B and IV.C of \cite{nik:08} and also in \cite{sey:nik:alb:12}, even if Alice publicly announces $n$ the system is still secure. In fact, using the Holevo bound, the cryptographic scheme is information secure if $k$, the length of the message is of same order than $n$. In fact, the maximum information that Bob can obtain from the public key regarding the secret key is $k$ bits, but its uncertainty is $nk$, which in case of $n=k$ implies a negligible advantage for an adversary to recover the message. Furthermore, since any rotation leaves a complete mixed state invariant, all possible messages yield the same cipher state, which is equal to a public-key state, and therefore adversaries cannot distinguish distinct messages, unless many copies of the encrypted state and the respective public key are provided, making the scheme provably secure~\cite{boy:roy:03,zou:qui:13}. 

\begin{figure}[H]
\begin{center}
\includegraphics[angle=0, width=11cm, keepaspectratio=true] 
{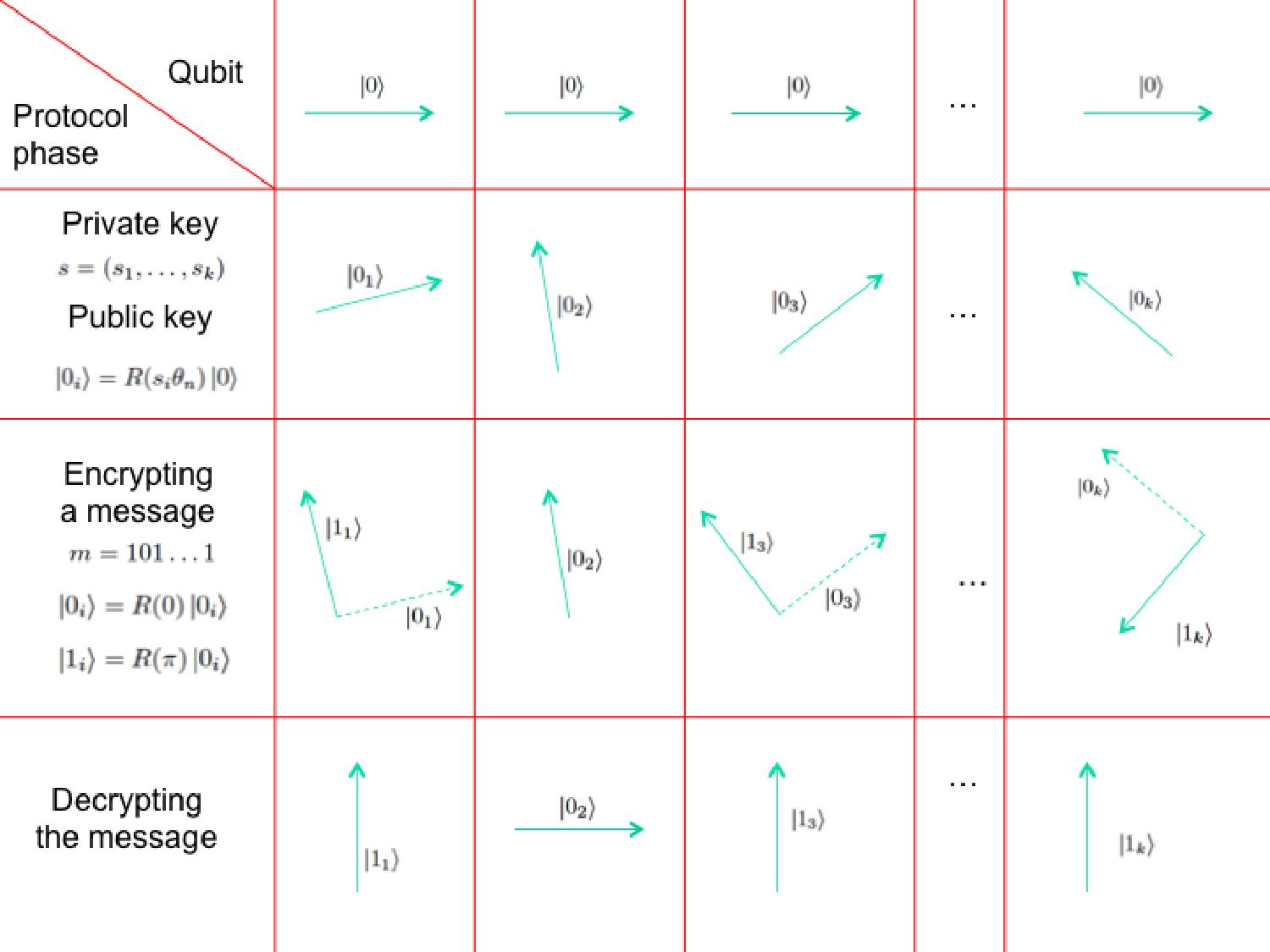}
\caption{Schematic description of the public key cryptosystem for messages of length $k$. The full arrows represent the actual states of qubits, while the dashed arrows in the third line (encryption of a message) represent $\ket{0_i}$ states.}
\label{fig:publickey}
\end{center}
\end{figure}

%
%Therefore Protocol~\ref{prot:PK} is secure.
In Figure~\ref{fig:publickey} we present a schematic description of the public-key cryptosystem.%For later reference we formally state the result as a theorem:
%
%\begin{theorem}\label{theo:security}
%Protocol \ref{prot:PK} is secure against plaintext recovery by any Eavesdropper. 
%\end{theorem}

At the end of the Oblivious transfer protocol, Bob has to be assured if he received the message or not. 
There are different ways to guarantee this feature. The solution adopted in this paper is to use a {\it hash function}.
A hash function maps strings to other strings of smaller size. Therefore, different strings are mapped to the same hash value. 
Hash functions have to satisfy the following two constrains,
\begin{itemize}
	\item their value for each input can be computed in polynomial time on the length of the input string;
	\item The hash values of a randomly chosen string are uniformly distributed.
 \end{itemize}
%and are defined as follows.
%\begin{definition}

	Consider two sets $A$ and $B$ of size $a$ and $b$, respectively, such that $a>b$, and consider a hash function $h:A\to B$.
%\end{definition} 
It is easy to derive that 
%and are defined as follows.
%\begin{definition}
%
%	Consider two sets $A$ and $B$ of size $a$ and $b$, respectively, such that 
%		$a>b$, and consider a collection $ H$ of hash functions $h:A\to B$.
%	If 
%	\begin{eqnarray}
%	\Pr_{h\in  H}[h(x) = h(y)] \leq\frac 1 b
%	\end{eqnarray}
%	then $ H$ is called a \emph{universal family of hash functions}.
%%\end{definition} 
%From the above definition, it is easy to derive that % if the following result:
%\begin{theorem}\label{theo:universalhash}
	%Let  $ H$ be a collection of hash functions $h:A\to B$, where $A$ and $B$ are sets of size $a$ and $b$, respectively, 
	%such that $a>b$. 
	%
	the size of a set $A_\mathbf{x}$ of strings $\mathbf{x}\in A$ mapped to the same hash value $h(\mathbf{x})$ is at most $a/b$.
%\end{theorem}

In particular, requiring that $A$ contains all strings of length $k$ and $B$ to be a set of strings of length $\omega = \lfloor \sqrt k\rfloor$, the number of strings with the same hash value is $2^{\omega}$, hence the probability of finding such a string is negligible in $k$. 
For more details on constructing hash functions, see for example~\cite{car:weg:79}.

%\section*{References}

\bibliography{quantum_andre}{}
\bibliographystyle{alpha}

\end{document}